\documentclass[aps, pre, twocolumn, floatfix]{revtex4}
\usepackage{graphicx}
\usepackage{amssymb, amsfonts, amsmath}
\usepackage{epsf}
\usepackage{subfigure}
\usepackage{epstopdf}
\usepackage{mathtools}
\usepackage{txfonts}
\usepackage{graphics}
\usepackage{float}
\usepackage{mathrsfs}
\usepackage{chemarrow}
\usepackage{bigstrut}
\usepackage{array}

\usepackage{color}

\newcommand{\be}{\begin{equation}}
\newcommand{\ee}{\end{equation}}
\newcommand{\bea}{\begin{eqnarray}}
\newcommand{\eea}{\end{eqnarray}}

\begin{document}

\title{Microreversibility, fluctuations, and nonlinear transport in transistors}

\author{Jiayin Gu}
\email{jiaygu@ulb.ac.be}
\author{Pierre Gaspard}
\email{gaspard@ulb.ac.be}
\affiliation{Center for Nonlinear Phenomena and Complex Systems, Universit\'e Libre de Bruxelles (U.L.B.), Campus Plaine, Code Postal 231, B-1050 Brussels, Belgium}

\begin{abstract}
We present a stochastic approach for charge transport in transistors. In this approach, the electron and hole densities are governed by diffusion-reaction stochastic differential equations satisfying local detailed balance and the electric field is determined with the Poisson equation. The approach is consistent with the laws of electricity, thermodynamics, and microreversibility.  In this way, the signal amplifying effect of transistors is verified under their working conditions. We also perform the full counting statistics of the two electric currents coupled together in transistors and we show that the fluctuation theorem holds for their joint probability distribution.  Similar results are obtained including the displacement currents.  In addition, the Onsager reciprocal relations and their generalizations to nonlinear transport properties deduced from the fluctuation theorem are numerically shown to be satisfied.
\end{abstract}

\maketitle

\section{Introduction}

Transistors are the main compounds of semiconductor electronic technology. The core of transistors is composed of three semiconducting materials concatenated in series, thus forming double junctions. The middle semiconductor is doped with charged impurities different from those in the two other semiconductors. Since transistors have three ports and currents flow between pairs of ports, two electric currents are coupled together inside transistors, enabling the amplification of signals \cite{SST51,EM54,SS04,CC05,B05,SN07}.

The fundamental issue is that the coupling between the electric currents is ruled by microreversibility, as in any type of device or process.  In linear regimes close to thermodynamic equilibrium, microreversibility implies the Onsager-Casimir reciprocal relations~\cite{O31a,O31b,C45}.  However, transistors are functioning in highly nonlinear regimes beyond the domain of application of the Onsager-Casimir reciprocal relations.  Remarkably, the generalizations of these relations beyond the linear regime are known today \cite{S92,AG04,AG07JSM,HPPG11,BG18}.  They can be deduced from the fluctuation theorem for currents, which is based on the time-reversal symmetry of the microscopic dynamics of electrons and ions \cite{AG07JSP,AGMT09,AG09,EHM09,CHT11,S12,G13}. The fluctuation theorem is valid not only in the linear regimes, but also in the nonlinear regimes, and can thus be used to investigate the nonlinear transport properties of transistors.

In our previous paper \cite{GG18}, the fluctuation theorem was considered for diodes that are also nonlinear electronic devices.  Here, our purpose is to extend these considerations to transistors.  The novel aspect is that two currents are flowing in transistors, instead of only one in diodes.  As a consequence of the nonlinear coupling between the two currents, the generalizations of Onsager-Casimir reciprocal relations to nonlinear transport can be tested in transistors.

For this purpose, the stochastic approach of Ref.~\cite{GG18} is extended from the single junction of diodes to the double junction of $n$-$p$-$n$ transistors.  The approach is based on diffusion-reaction stochastic partial differential equations for electrons and holes, including their Coulomb interaction described by the Poisson equation.  This scheme satisfies local detailed balance in consistency with microreversibility.  The stochastic description is presented in Sec.~\ref{sec:stochastic}.  The functionality of transistors is studied in Sec.~\ref{sec:funct}.  Section~\ref{sec:FT} is devoted to the fluctuation theorem for the two currents of the transistor.  Section~\ref{sec:resp} shows that the linear response coefficients obey the Onsager-Casimir reciprocal relation and the fluctuation-dissipation theorem, and that the next-order nonlinear response coefficients satisfy higher-order generalizations.  Section~\ref{sec:conclude} gives concluding remarks.

\section{Stochastic description of transistors}
\label{sec:stochastic}

\subsection{The bipolar $n$-$p$-$n$ junction transistor}

\begin{figure*}
\begin{minipage}[t]{0.28\hsize}
\resizebox{1.0\hsize}{!}{\includegraphics{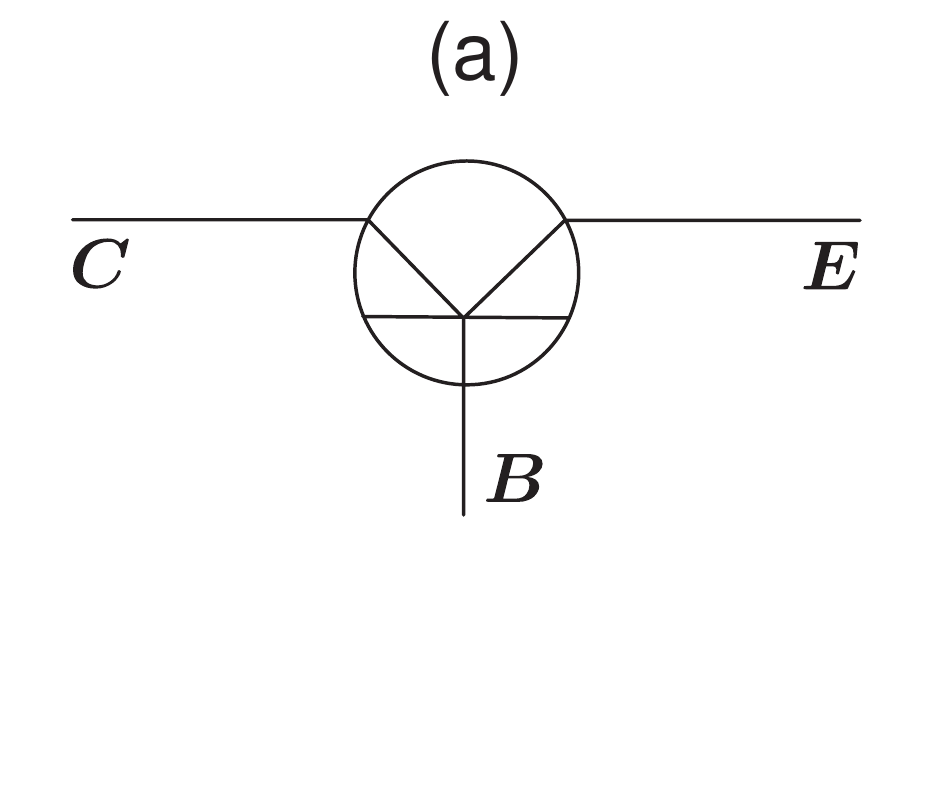}}
\end{minipage}
\begin{minipage}[t]{0.50\hsize}
\resizebox{1.0\hsize}{!}{\includegraphics{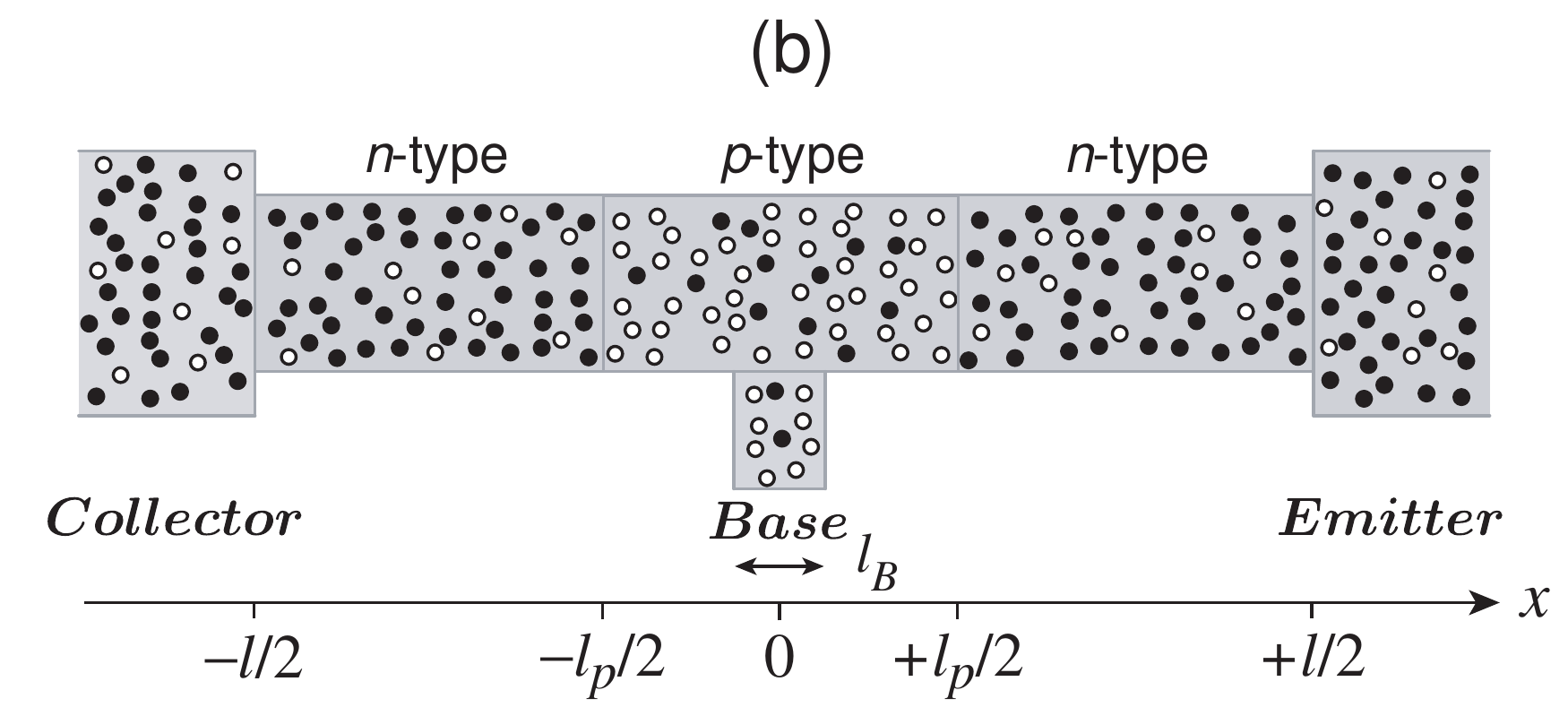}}
\end{minipage}
\caption{Schematic representation of (a) the transistor and (b) the bipolar $n$-$p$-$n$ double junction. In panel~(b), the black (resp. white) dots represent electrons (resp. holes). The three reservoirs, called \textit{Collector}, \textit{Base}, and \textit{Emitter}, fix the values of the electron density, the hole density, and the electric potentials at their contact with the transistor.}
\label{fig1}
\end{figure*}

There exists many types of transistors \cite{SS04,CC05,B05,SN07}. The bipolar $n$-$p$-$n$ junction transistor (BJT) is one of the most common of them. BJTs consist of three small doped regions of a piece of silicon, respectively typed as $n$, $p$, and $n$, thus forming two junctions, as shown in Fig.~\ref{fig1}. The electrons~${\rm e}^{-}$ and holes~${\rm h}^{+}$ are the two mobile charge carriers across the bipolar $n$-$p$-$n$ junction, with electrons being the majority ones in $n$-type semiconductor, and holes the majority ones in $p$-type semiconductor. The positively-charged donors and negatively-charged acceptors are respectively anchored in $n$-type semiconductors and $p$-type semiconductors. Each doped region has a port and the three ports are in contact with some charge carrier reservoir. They are respectively called \textit{Collector}, \textit{Base}, and \textit{Emitter} (see Fig.~\ref{fig1}).   

In order to model the transistor, a Cartesian coordinate system is associated with the system.  As shown in Fig.~\ref{fig1}(b), the semiconducting material extends from $x=-l/2$ to $x=+l/2$ and is divided in three parts.  The part from $x=-l/2$ to $x=-l_p/2$ is of $n$-type, the one from $x=-l_p/2$ to $x=+l_p/2$ of $p$-type, and the one from $x=+l_p/2$ to $x=+l/2$ of $n$-type.  The three parts are respectively of lengths $l_n=(l-l_p)/2$, $l_p$, and $l_n=(l-l_p)/2$.  The {\it Collector} is in contact at $x=-l/2$, the {\it Emitter} at $x=+l/2$, and the {\it Base} along a length $l_B$ symmetrically located around the origin $x=0$.  The length of the contact with the {\it Base} is smaller than the one of the $p$-type part: $l_B < l_p$.  The geometry is chosen to be symmetric with respect to $x=0$ for simplicity.

In addition, the bipolar $n$-$p$-$n$ double junction has the section area $\Sigma$ in the transverse $y$- and $z$-directions. The section areas of the contacts with the {\it Collector} and {\it Emitter} are assumed to be equal: $\Sigma_C=\Sigma_E=\Sigma$.  Accordingly, the semiconducting material extends over a domain of volume $V=l\Sigma$.  Moreover, we denote $\Sigma_B$ the section area of the contact with the {\it Base}.  

The donor density $d({\bf r})$ and acceptor density $a({\bf r})$ are supposed to be uniform in the different types of semiconductor. Therefore, they can be expressed as
\begin{align}
& d({\bf r})=d\, \theta\left(-x-l_p/2\right)+d\, \theta\left(x-l_p/2\right) \text{,} \\
& a({\bf r})=a\, \theta\left(x+l_p/2\right)\, \theta\left(-x+l_p/2\right) \text{,}
\end{align}
in terms of two constant values $a$ and $d$, combined with Heaviside's step function $\theta(x)$ defined such that $\theta(x)=1$ if $x>0$ and $\theta(x)=0$ otherwise. The charge density is thus given by
\begin{align}
\rho=e(p-n+d-a) \, , \label{eq_charge_density}
\end{align}
with the elementary electric charge $e=|e|$, and the densities of holes $p$, electrons $n$, donors $d$, and acceptors $a$.  Here, we have assumed that every donor gives one electron and every acceptor one hole.  Because of the electrostatic interaction between the charges, these densities are coupled to the electric potential $\phi({\bf r})$.

The electron and hole densities as well as the electric potential have fixed boundary values at the contacts with the three reservoirs.  They are respectively given by $n_C$, $p_C$, $\phi_C$ at the {\it Collector}; $n_B$, $p_B$, $\phi_B$ at the {\it Base}; and $n_E$, $p_E$, $\phi_E$ at the {\it Emitter}.  

If the transistor is at equilibrium without flow of charge carriers, detailed balance between the generation and recombination of electron-hole pairs requires that $n_{\rm eq} p_{\rm eq}=\nu^2$, where $\nu$ is called the intrinsic carrier density.  Moreover, the electron and hole densities are given at equilibrium by 
\begin{align}
n_{\rm eq}({\bf r})\sim {\rm e}^{+\beta\phi_{\rm eq}({\bf r})}\hspace{1cm}\text{and}\hspace{1cm} p_{\rm eq}({\bf r})\sim{\rm e}^{-\beta e\phi_{\rm eq}({\bf r})}
\end{align}
in terms of the electric potential determined across the whole system by the Poisson equation and the boundary conditions at the contacts with the three reservoirs. If the BJT is at equilibrium, the inhomogeneous distributions of the charge carriers thus produce the Nernst potentials
\begin{align}\label{Nernst_C_E}
(\phi_C-\phi_E)_{\rm eq}=\frac{1}{\beta e}\ln\frac{n_C}{n_E}=\frac{1}{\beta e}\ln\frac{p_E}{p_C}
\end{align}
and
\begin{align}\label{Nernst_B_E}
(\phi_B-\phi_E)_{\rm eq}=\frac{1}{\beta e}\ln\frac{n_B}{n_E}=\frac{1}{\beta e}\ln\frac{p_E}{p_B} \text{,}
\end{align}
where $\beta\equiv(k_{\rm B}T)^{-1}$ is the inverse temperature.  

The transistor is driven out of equilibrium by applying voltage differences with respect to the Nernst potentials
\begin{align}
& V_C=\phi_C-\phi_E-\frac{1}{\beta e}\ln\frac{n_C}{n_E} \text{,} \label{eq_V_C}\\
& V_B=\phi_B-\phi_E-\frac{1}{\beta e}\ln\frac{n_B}{n_E} \text{,} \label{eq_V_B}
\end{align}
which induce currents across the BJT. In the following, we use the associated affinities or thermodynamic forces
\begin{align}\label{affinities}
A_C \equiv \beta e V_C \qquad\mbox{and} \qquad A_B \equiv \beta e V_B\text{,}
\end{align}
which are dimensionless. The equilibrium state is recovered if they vanish, i.e., if the applied voltages are equal to zero $V_C=V_B=0$.

\subsection{Stochastic diffusion-reaction equations}

The thermal agitation inside the BJT generates incessant erratic motion for the electrons and holes, in turn causing local fluctuations in the currents and reaction rates. These fluctuations can be described within the stochastic approach by introducing Gaussian white noise fields in the diffusion-reaction equations for the electron and hole densities.  The advantage of this approach is that the usual phenomenological parameters suffice for the stochastic description.

The mobilities of electrons and holes are related with their diffusion coefficients through Einstein's relations
\begin{align}
\mu_n=\beta e D_n \hspace{1cm}\text{and}\hspace{1cm} \mu_p=\beta eD_p \text{.}
\end{align}
Besides, the electron-hole pairs are randomly generated and recombined according to the reactions
\begin{align}
{\rm e}^{-}+{\rm h}^{+}\autorightleftharpoons{$\scriptstyle k_-$}{$\scriptstyle k_+$}\emptyset  \, \text{,}
\label{eq_reaction}
\end{align}
where $k_{+}$ and $k_{-}$ are respectively the generation and recombination rate constants. In general, the quantities $D_n$, $D_p$, and $k_{\pm}$ are spatially dependent in an inhomogeneous medium. However, for simplicity, we assume that they are uniform across the whole BJT.

Considering the diffusion and generation-recombination processes as well as the electrostatic interaction between the charges, we have the following stochastic partial differential equations for the charge carrier densities coupled to the Poisson equation for the electric potential,
\begin{align}
& \partial_tn+{\bf\nabla}\cdot{\bf j}_n=\sigma_n \text{,} \label{eq-n}\\
& \partial_tp+{\bf\nabla}\cdot{\bf j}_p=\sigma_p \text{,} \label{eq-p}\\
& \nabla^2\phi=-\frac{\rho}{\epsilon} \text{,} \label{eq-phi}
\end{align}
where 
\begin{align}
& \sigma_n=\sigma_p=k_+-k_-np+\delta\sigma \text{,} \label{eq-s}\\
& {\bf j}_n=-\mu_n n\,\pmb{\cal E}-D_n{\bf\nabla}n+\delta{\bf j}_n \text{,} \label{eq-jn}\\
& {\bf j}_p=+\mu_p p\,\pmb{\cal E}-D_p{\bf\nabla}p+\delta{\bf j}_p \text{,} \label{eq-jp}\\
& \pmb{\cal E}=-{\bf\nabla}\phi \text{,} \label{eq-E}
\end{align}
are the reaction rates, the current densities, and the electric field, while $\rho$ is the charge density given by Eq. (\ref{eq_charge_density}) and $\epsilon$ the dielectric constant of the material \cite{GG18}.  The fluctuations $\delta{\bf j}_n$, $\delta{\bf j}_p$, and $\delta\sigma$ are Gaussian white noise fields characterized by
\begin{align}
& \langle \delta{\bf j}_n({\bf r},t) \rangle = \langle \delta{\bf j}_p({\bf r},t) \rangle = 0 \label{av_j}\text{,} \\
& \langle \delta{\bf j}_n({\bf r},t)\otimes \delta{\bf j}_n({\bf r}',t') \rangle = \Gamma_{nn}({\bf r},t) \, \delta^3({\bf r}-{\bf r'}) \, \delta(t-t') \, {\boldsymbol{\mathsf 1}} \text{,} \\
& \langle \delta{\bf j}_p({\bf r},t)\otimes \delta{\bf j}_p({\bf r}',t') \rangle = \Gamma_{pp}({\bf r},t) \, \delta^3({\bf r}-{\bf r'}) \, \delta(t-t') \, {\boldsymbol{\mathsf 1}} \text{,} \\
& \langle \delta{\bf j}_n({\bf r},t)\otimes \delta{\bf j}_p({\bf r}',t') \rangle = 0 \text{,} \\
& \langle\delta\sigma({\bf r},t)\rangle = 0 \text{,} \label{av_s}\\
& \langle\delta\sigma({\bf r},t)\,\delta\sigma({\bf r'},t')\rangle = \Gamma_{\sigma\sigma}({\bf r},t) \, \delta^3({\bf r}-{\bf r'}) \, \delta(t-t') \text{,}  \\
& \langle \delta\sigma({\bf r},t)\, \delta{\bf j}_n({\bf r}',t') \rangle = \langle \delta\sigma({\bf r},t)\, \delta{\bf j}_p({\bf r}',t') \rangle = 0 \text{,}
\end{align}
where ${\boldsymbol{\mathsf 1}}$ is the $3\times 3$ identity matrix and
\begin{align}
&\Gamma_{nn}({\bf r},t) \equiv 2\, D_n \, n({\bf r},t) \text{,} \label{GWN-n}\\
&\Gamma_{pp}({\bf r},t) \equiv 2\, D_p \, p({\bf r},t) \text{,} \label{GWN-p}\\
&\Gamma_{\sigma\sigma}({\bf r},t) \equiv k_++k_- n({\bf r},t) p({\bf r},t) \label{GWN-sig}
\end{align}
are the noise spectral densities associated with the electron and hole diffusions, and the reaction.

Because of Eqs.~(\ref{av_j}) and~(\ref{av_s}), we recover the mean-field equations of the macroscopic description by averaging the stochastic partial differential equations over the noises.

\subsection{Numerical method for simulating the transistor}

For the numerical simulation of the transistor, a Markov jump process is associated with the stochastic partial differential equations~(\ref{eq-n})-(\ref{eq-E}), as described in detail in Appendix~\ref{App:Markov}.  Space is discretized into $L$ cells of length $\Delta x=l/L$, section area $\Sigma$, and volume $\Omega=\Sigma\Delta x$, located at the coordinates $x_{i}=(i-0.5)\Delta x-l/2$ ($i=1,2,\dots,L$).  Consistently with Fig.~\ref{fig1}(b), there are $L_n=l_n/\Delta x$ cells in both parts of $n$-type, $L_p=l_p/\Delta x$ cells for the part of $p$-type, and $L_B=l_B/\Delta x$ cells in contact with the {\it Base}.  The numbers of electrons, holes, acceptors, and donors in each cell of the BJT are related to the corresponding densities by $N_{i}=n(x_{i})\Omega$, $P_{i}=p(x_{i})\Omega$, $A_{i}=a(x_{i})\Omega$, and $D_{i}=d(x_{i})\Omega$.  The state of the discretized BJT is fully characterized by the electron numbers ${\bf N}=(N_i)_{i=1}^L$ and the hole numbers ${\bf P}=(P_i)_{i=1}^L$ in the cells.  The master equation ruling the time evolution of their probability distribution ${\cal P}({\bf N},{\bf P},t)$ is given in Appendix~\ref{master_eq}.  Moreover, the Poisson equation~(\ref{eq-phi}) is also discretized along the chain of $L$ cells forming the system,  taking into account the electric potentials of the {\it Collector}, the {\it Base}, and the {\it Emitter}, as explained in Appendix~\ref{App:Poisson}.  The resulting stochastic process can be simulated numerically by Gillespie's algorithm \cite{G76}, which is an exact method for generating random trajectories in this case.

In order to speed up the simulation, the Markov jump process is approximated by a Langevin stochastic process under the assumption that the numbers of electrons and holes are large enough in every cell, $N_i\gg 1$ and $P_i\gg 1$.  Accordingly, these numbers obey stochastic differential equations expressed in terms of the fluxes of particles between the cells, the reaction rates, and Gaussian white noises for their fluctuations, as shown in Appendix \ref{App:Langevin}.

At the contacts with the three reservoirs, the boundary conditions on the charge carrier densities determine the boundary values for the corresponding particle numbers
\begin{align}
& \bar{N}_C=n_C\Omega\text{,}\hspace{1cm} \bar{P}_C=p_C\Omega \text{,}\\
& \bar{N}_B=n_B\Omega\text{,}\hspace{1cm} \bar{P}_B=p_B\Omega \text{,}\\
& \bar{N}_E=n_E\Omega\text{,}\hspace{1cm} \bar{P}_E=p_E\Omega \text{.}
\end{align}
Furthermore, the three parts of the transistor are supposed to be doped from a semiconducting material of uniform  intrinsic density $\nu$, so that the boundary values of the electron and hole densities should satisfy the conditions
\begin{align}
n_Cp_C=n_Bp_B=n_Ep_E = \nu^2 \text{.}
\end{align}
We further set
\begin{align}
n_C=n_E\text{,}\hspace{1cm}p_C=p_E \text{,}
\end{align}
to have a system that is symmetric with respect to $x=0$, as depicted in Fig.~\ref{fig1}(b).  

In numerical simulations, the statistical averages of any observable quantity $X$ can be evaluated by the time average $\langle X\rangle=\lim_{T\to\infty}(1/T) \int_0^TX(t)\, dt$, which is equivalent by ergodicity  to the ensemble average $\langle X\rangle=\sum_{{\bf N},{\bf P}}X\,{\cal P}_{\rm st}({\bf N},{\bf P})$ over the stationary probability distribution ${\cal P}_{\rm st}$.  In the continuum limit, the volume of the cells is supposed to vanish together with the particle numbers, so that the electron and hole densities can be recovered as $n(x_{i})=N_{i}/\Omega$ and $p(x_{i})=P_{i}/\Omega$.

\bigstrutjot=2pt
\begin{table*}[h]
\caption{The values of dimensionless physical quantities and parameters used in simulating the BJT model in rescaled units.}
\begin{tabular}{>{\centering\arraybackslash}m{6cm}>{\centering\arraybackslash}m{2cm}||>{\centering\arraybackslash}m{6cm}>{\centering\arraybackslash}m{2cm}}
\hline
\hline
quantity & value & quantity & value \bigstrut \\ \hline
permittivity   & $\epsilon=0.01$ &  length of each cell   & $\Delta x=0.1$ \bigstrut \\ \hline
elementary charge & $|e|=1.0$ & width of each cell & $\Delta y=0.2$ \bigstrut \\ \hline
inverse temperature & $\beta=1.0$ & number of cells in both $n$-type regions & $L_n=10$  \bigstrut \\ \hline
diffusion coefficient for electrons and holes & $D=0.01$ & number of cells in the $p$-type region & $L_p=3$ \bigstrut \\ \hline
generation and recombination rate constants & $k_+=k_-=0.01$ & number of cells in contact with the {\it Base} & $L_B=1$ \bigstrut \\ \hline
\hline
\end{tabular}
\label{tab_physical_quantities}
\end{table*}

\begin{table*}[h]
\caption{The set of parameter values used in Sec.~\ref{sec:funct}.}
\begin{tabular}{>{\centering\arraybackslash}m{5cm}>{\centering\arraybackslash}m{3cm}||>{\centering\arraybackslash}m{5cm}>{\centering\arraybackslash}m{3cm}}
\hline
\hline
parameter & value & parameter & value \bigstrut \\ \hline
volume of each cell   & $\Omega=10^9$ & section area & $\Sigma=10^{10}$, $\Sigma_B=5\times 10^9$ \bigstrut \\ \hline
number of electrons for the \textit{Collector} & $\bar{N}_C=10^{13}$ & number of holes for the \textit{Collector} & $\bar{P}_C=10^5$ \bigstrut \\ \hline
number of electrons for the \textit{Base} & $\bar{N}_B=10^8$ & number of holes for the \textit{Base} & $\bar{P}_B=10^{10}$ \bigstrut \\ \hline
number of electrons for the \textit{Emitter} & $\bar{N}_E=10^{13}$ & number of holes for the \textit{Emitter} & $\bar{P}_E=10^5$ \bigstrut \\ \hline
\hline
\end{tabular}
\label{tab_parameters_1}
\end{table*}

\begin{figure*}[h]
\begin{minipage}[t]{0.99\hsize}
\resizebox{1.0\hsize}{!}{\includegraphics{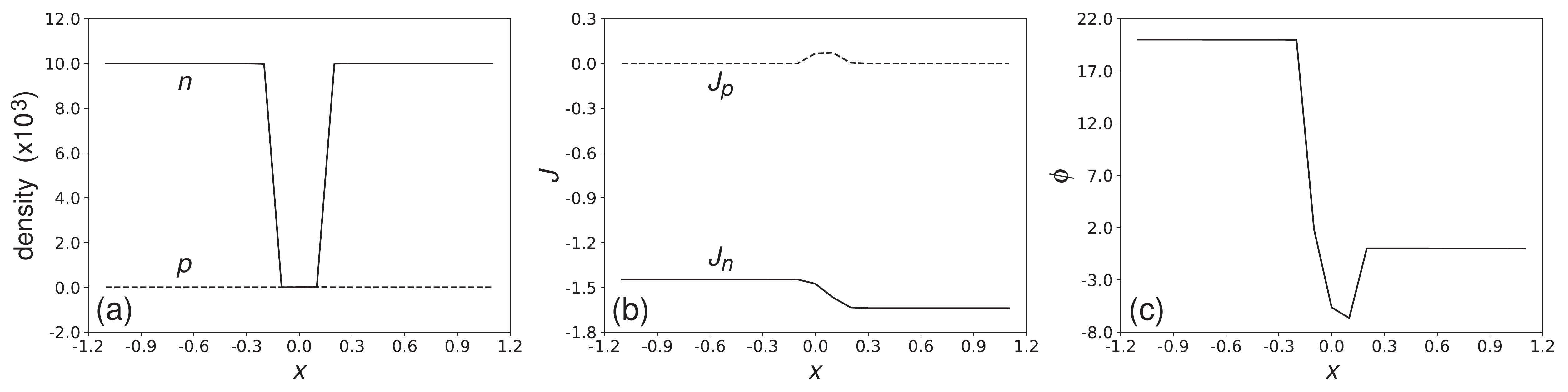}}
\end{minipage}
\caption{The profiles of (a) the charge carrier densities, (b) the current densities, and (c) the electric potential across the BJT which is used as signal amplifier under the working conditions $A_C=20$ and $A_B=6$. The {\it Collector C} is located at $x\leq-1.15$, the {\it Emitter E} at $x\geq+1.15$, and the {\it Base B} around $x=0$.  The simulations were carried out with the time step $dt=0.00015$ and $10^6$ iterates for every data point.}
\label{fig2}
\end{figure*}

\begin{figure*}[h]
\begin{minipage}[t]{0.99\hsize}
\resizebox{1.0\hsize}{!}{\includegraphics{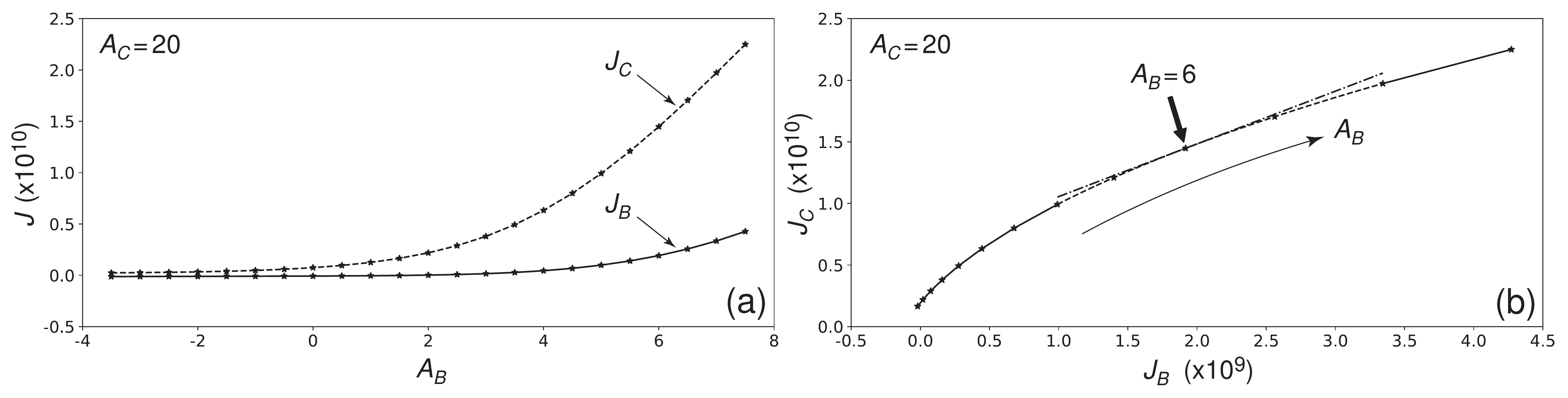}}
\end{minipage}
\caption{(a) The mean currents $J_C$ and $J_B$ versus the affinity $A_B$, with the other affinity fixed to the value $A_C= 20$. The lines join the numerical points depicted by the asterisks. (b) The current $J_C$ versus the other current $J_B$. The solid line joins the asterisks. The dashed line in the middle region is determined from Lagrange interpolation using the five asterisks of this domain. The derivative of $J_C$ with respect to $J_B$ at the point $(A_C=20,A_B=6)$ is evaluated giving the amplification factor~(\ref{alpha_num}).  The simulations were carried out with the time step $dt=0.00015$ and $10^6$ iterates for every data point.}
\label{fig3}
\end{figure*}

We assume for simplicity that the electron and hole diffusion coefficients are equal $D_n=D_p\equiv D$.  As done in our previous paper~\cite{GG18}, the quantities of interest may be rescaled using the intrinsic carrier density $\nu$, the intrinsic carrier lifetime $\tau=1/(k_-\nu)$, the intrinsic carrier diffusion length before recombination $\ell=\sqrt{D\tau}$, the inverse temperature $\beta$, and the elementary electric charge.  After this rescaling, the quantities of interest become dimensionless.  Table~\ref{tab_physical_quantities} gives the values of the so-rescaled quantities used in the following numerical simulations of the BJT model.

\section{The Functionality of transistors}
\label{sec:funct}

The purpose of this section is to show that the properties characterizing the functionality of transistors can be described within the stochastic approach. 

In electronic technology, transistors are primarily used to amplify signals in electric circuits. This amplification results from the coupling between the two electric currents, $J_C$ and $J_B$. By this coupling, one current can serve as input and the other as output. The amplification factor is defined as the ratio of these two currents, $J_C/J_B$.  We may also introduce the differential amplification factor as follows.  When the affinity $A_C$ is fixed, the variation of the other affinity $A_B$ leads to variations of $J_C$ and $J_B$. The amplification factor  is defined as the ratio between these two variations
\begin{align}
\alpha=\left(\frac{\partial J_C}{\partial J_B}\right)_{A_C}
\label{alpha-dfn}
\end{align}
under specific working conditions.  To achieve the functionality of signal amplification, the transistors should satisfy the following requirements:
\begin{itemize}
\item The concentration of the majority charge carriers in the \textit{Collector} region should be overwhelmingly larger than the concentration of minority charge carriers in the \textit{Base} region.
\item The concentration of the majority charge carriers in the \textit{Emitter} region should be overwhelmingly larger than the concentration of minority charge carriers in the \textit{Base} region.
\item The \textit{Collector}-\textit{Base} junction should be reverse biased.
\item The \textit{Emitter}-\textit{Base} junction should be forward biased.
\item The \textit{Base} region should be very thin so that the majority charge carriers in the \textit{Emitter} region can easily get swept to the \textit{Collector} region.
\item The contacting section areas $\Sigma_C$ and $\Sigma_E$ should be larger than $\Sigma_B$.
\end{itemize}

Table \ref{tab_parameters_1} gives a set of parameter values approaching these requirements in order to show that the present stochastic model can describe transistors in such regimes.  The first two conditions are satisfied since $\bar{N}_C=\bar{N}_E\gg\bar{N}_B$, and the last one because $\Sigma=\Sigma_C=\Sigma_E>\Sigma_B$.

If the transistor was at equilibrium without applied voltage ($A_C= A_B=0$), the Nernst potentials~(\ref{Nernst_C_E}) and~(\ref{Nernst_B_E}) would take the values $(\phi_C-\phi_E)_{\rm eq}=0$ and $(\phi_B-\phi_E)_{\rm eq}=-11.5$ with the parameter set of Table \ref{tab_parameters_1}.  At equilibrium, the electric field would have a symmetric profile around $x=0$ with $(\phi_C-\phi_B)_{\rm eq}=(\phi_E-\phi_B)_{\rm eq}=11.5$.

Figure~\ref{fig2} shows the profiles of charge carrier densities and current densities together with the electric potential under nonequilibrium conditions with applied voltages corresponding to $A_C= 20$ and $A_B=6$.
In Fig.~\ref{fig2}(a), we see that the {\it Base} region is thin in the model, so that the fifth condition is satisfied.  
As observed in Fig.~\ref{fig2}(b), the current densities are non-vanishing because the transistor is out of equilibrium.
According to Eqs.~(\ref{eq_V_C})-(\ref{eq_V_B}), we here have that $\phi_C-\phi_E=20$ and $\phi_B-\phi_E=-5.5$, so that $\phi_C-\phi_B=25.5$ and $\phi_E-\phi_B=5.5$, in agreement with the electric field plotted in Fig.~\ref{fig2}(c).  Since $\phi_C-\phi_B=25.5$ is larger than $(\phi_C-\phi_B)_{\rm eq}=11.5$, the \textit{Collector}-\textit{Base} junction is reverse biased, as it should by the third condition.  Moreover, $\phi_E-\phi_B=5.5$ is smaller than $(\phi_E-\phi_B)_{\rm eq}=11.5$, so that the \textit{Emitter}-\textit{Base} junction is forward biased and the fourth condition is also satisfied.  Under these conditions, the transistor can indeed achieve signal amplification, as demonstrated in Fig.~\ref{fig3}.  The currents $J_C$ and $J_B$ are shown in Fig.~\ref{fig3}(a) as functions of $A_B$, with $A_C$ fixed.  Since the current $J_C$ is greater than $J_B$, the amplification factor $J_C/J_B$ is larger than unity, as expected.  Furthermore, Fig.~\ref{fig3}(b) depicts how the current $J_C$ increases with the other current $J_B$ and the associated affinity $A_B$.  For $A_B=6$, the differential amplification factor~(\ref{alpha-dfn}) is evaluated to be
\begin{align}
\alpha(A_C= 20,A_B=6)\simeq 4.278 \text{,}
\label{alpha_num}
\end{align}
which is also larger than unity, as required.   It should be noticed that the amplification factors can take different values for different working conditions of the transistor.

These results show that the stochastic approach is relevant to study transistors in their regimes of signal amplification. We proceed in the next Sec.~\ref{sec:FT} and Sec.~\ref{sec:resp} with the study of their fluctuation properties.

\section{Fluctuation Theorem for Currents}
\label{sec:FT}

\subsection{Generalities}

We consider the fluctuating electric currents flowing respectively across the contact with the \textit{Collector} and the contact with the \textit{Base}. These electric currents are due to the random motion of electrons and holes crossing the contact sections between the transistor and the corresponding reservoirs. The instantaneous electric currents are thus defined as
\begin{align}
&{\cal I}_C(t)=\sum_{n=-\infty}^{+\infty}q_n^{(C)}\delta(t-t_n^{(C)}) \text{,}\\
& {\cal I}_B(t)=\sum_{n=-\infty}^{+\infty}q_n^{(B)}\delta(t-t_n^{(B)}) \text{,}
\end{align}
where $t_n^{(C)}$ (resp. $t_n^{(B)}$) are the random times of the crossing events and $q_n^{(C)}$ (resp. $q_n^{(B)}$) are the transferred charges equal to $\pm e$ depending on whether the carrier is an electron or a hole and if its motion is inward or outward the transistor.  The corresponding random numbers of charges accumulated over the time interval $[0,\,t]$ are defined as
\begin{align}
Z_C(t)=\frac{1}{e}\int_0^{t}{\cal I}_C(t')\,dt' \text{,}\qquad Z_B(t)=\frac{1}{e}\int_0^{t}{\cal I}_B(t')\,dt' \, .
\label{Z-dfn}
\end{align}

We also define the instantaneous total electric currents including the contribution of displacement currents as
\begin{align}
&\tilde{\cal I}_C(t)={\cal I}_C(t)-\epsilon\, \partial_t\partial_x\phi\, \Sigma_C \text{,}\\ 
&\tilde{\cal I}_B(t)={\cal I}_B(t)-\epsilon\, \partial_t\partial_y\phi\, \Sigma_B \text{,}
\end{align}
which are the experimentally measured electric currents \cite{GG18,AG09,BB00,S38,R39}, as well the corresponding accumulated charge numbers $\tilde Z_C(t)$ and $\tilde Z_B(t)$ with definitions as in Eq.~(\ref{Z-dfn}).

The mean values of the charge currents are given by
\begin{align}
& J_C\equiv \lim_{t\to\infty}\frac{1}{t}\, \langle Z_C(t)\rangle = \lim_{t\to\infty}\frac{1}{t}\, \langle \tilde{Z}_C(t)\rangle\text{,} \label{J_C}\\
& J_B\equiv \lim_{t\to\infty}\frac{1}{t}\, \langle Z_B(t)\rangle = \lim_{t\to\infty}\frac{1}{t}\, \langle \tilde{Z}_B(t)\rangle\text{,} \label{J_B}
\end{align}
and the corresponding electric currents by $I_C=eJ_C$ and $I_B=eJ_B$.  The equality between the mean values without and with the displacement currents comes from the fact that the displacement currents are given by a time derivative.

The diffusivities of the currents are defined as
\begin{align}
&D_{CC}\equiv\lim_{t\to\infty}\frac{1}{2t}\, {\rm var}_{Z_CZ_C}(t)=\lim_{t\to\infty}\frac{1}{2t}\, {\rm var}_{\tilde Z_C\tilde Z_C}(t)\, , \label{D_CC}\\
&D_{BB}\equiv\lim_{t\to\infty}\frac{1}{2t}\, {\rm var}_{Z_BZ_B}(t)=\lim_{t\to\infty}\frac{1}{2t}\, {\rm var}_{\tilde Z_B\tilde Z_B}(t)\, , \label{D_BB}\\
&D_{CB}\equiv\lim_{t\to\infty}\frac{1}{2t}\, {\rm cov}_{Z_CZ_B}(t)=\lim_{t\to\infty}\frac{1}{2t}\, {\rm cov}_{\tilde Z_C\tilde Z_B}(t) \label{D_CB}
\end{align}
in terms of the variances and the covariances between the accumulated random charge numbers
\begin{align}
& {\rm var}_{Z_CZ_C}(t)\equiv\langle Z_C(t)Z_C(t)\rangle-\langle Z_C(t)\rangle^2 \text{,}\label{eq_var_CC} \\
& {\rm var}_{Z_BZ_B}(t)\equiv\langle Z_B(t)Z_B(t)\rangle-\langle Z_B(t)\rangle^2 \text{,}\label{eq_var_BB} \\
& {\rm cov}_{Z_CZ_B}(t)\equiv\langle Z_C(t)Z_B(t)\rangle-\langle Z_C(t)\rangle\langle Z_B(t)\rangle = {\rm cov}_{Z_BZ_C}(t)\text{.}\label{eq_cov_CB}
\end{align}
The diffusivities also take the same value whether the displacement currents are included or not.  Since the covariance between two random variables is symmetric under their exchange, we have the symmetry $D_{CB}=D_{BC}$.

We suppose that the voltages~(\ref{eq_V_C}) and~(\ref{eq_V_B}) are applied at the boundaries of the transistor.  Consequently, the transistor is driven out of equilibrium and the stochastic process of charge transfers between the reservoirs eventually reaches a nonequilibrium steady state.  This latter is expected to depend on the applied voltages, or equivalently on the affinities
\begin{align}
& A_C=\ln\left[\frac{\bar{P}_C}{\bar{P}_E}{\rm e}^{\beta e(\phi_C-\phi_E)}\right]=\ln\left[\frac{\bar{N}_E}{\bar{N}_C}{\rm e}^{\beta e(\phi_C-\phi_E)}\right]=\beta eV_C \text{,} \label{eq_theoretical_affinity_C} \\
& A_B=\ln\left[\frac{\bar{P}_B}{\bar{P}_E}{\rm e}^{\beta e(\phi_B-\phi_E)}\right]=\ln\left[\frac{\bar{N}_E}{\bar{N}_B}{\rm e}^{\beta e(\phi_B-\phi_E)}\right]=\beta eV_B \text{,} \label{eq_theoretical_affinity_B}
\end{align} 
which are determined by the differences of electrochemical potentials between the corresponding reservoirs.  The dependences of the mean values of the currents on the affinities define the characteristic functions of the transistor: $J_C(A_C,A_B)$ and $J_B(A_C,A_B)$.  At equilibrium, the affinities are vanishing together with the applied voltages and the mean values of the currents, so that $J_C(0,0)=J_B(0,0)=0$.  However, the diffusivities do not necessarily vanish at equilibrium.

Beyond the mean values of the currents and the diffusivities, the process can be characterized by higher cumulants or the full probability distribution $P_{A_C,A_B}(Z_C,Z_B,t)$ that $Z_C(t)$ and $Z_B(t)$ charges are crossing the {\it Collector} and the {\it Base} during the time interval $[0,t]$, while the transistor is in a nonequilibrium steady state of affinities $A_C$ and $A_B$.  This steady state is given by the stationary solution of the master equation of the Markov jump process described in Appendix~\ref{App:Markov}.  Using the network representation of this Markov jump process and its decomposition into cyclic paths \cite{S76}, the process can be shown to obey a fluctuation theorem for all the currents as a consequence of local detailed balance \cite{AG07JSP,AG09}.  This theorem states that the joint distribution of random variables $Z_C$ and $Z_B$ at time $t$ satisfies the following fluctuation relation
\begin{align}
\frac{P_{A_C,A_B}(Z_C,Z_B,t)}{P_{A_C,A_B}(-Z_C,-Z_B,t)}\simeq_{t\to\infty}\exp(A_CZ_C+A_BZ_B) \text{.} \label{FT}
\end{align}
A similar fluctuation relation holds if the displacement currents are included in the accumulated charge numbers \cite{AG09}.

As a consequence of the fluctuation theorem, the thermodynamic entropy production is always non-negative in accord with the second law of thermodynamics.  The entropy production can indeed be expressed as the Kullback-Leibler divergence between the probability distributions of opposite fluctuations of the currents \cite{G13}, giving the dissipated power divided by the thermal energy
\be
\frac{1}{k_{\rm B}}\frac{d_{\rm i}S}{dt}= A_CJ_C+A_BJ_B=\beta \left( V_CI_C+V_BI_B\right) \geq 0 \, ,
\ee
as expected. 

We notice that the fluctuation relation~(\ref{FT}) holds in the long-time limit.
The convergence time is determined by diffusion~\cite{GGHK18} and it can be estimated to range between the time of diffusion across the middle part, $t_{\rm diff}\sim l_p^2/D\sim 9$, and the one before recombination, $t_{\rm diff}\sim \ell^2/D\sim 100$.

\subsection{Numerical results}

\begin{figure*}[h]
\begin{minipage}[t]{0.9\hsize}
\resizebox{1.0\hsize}{!}{\includegraphics{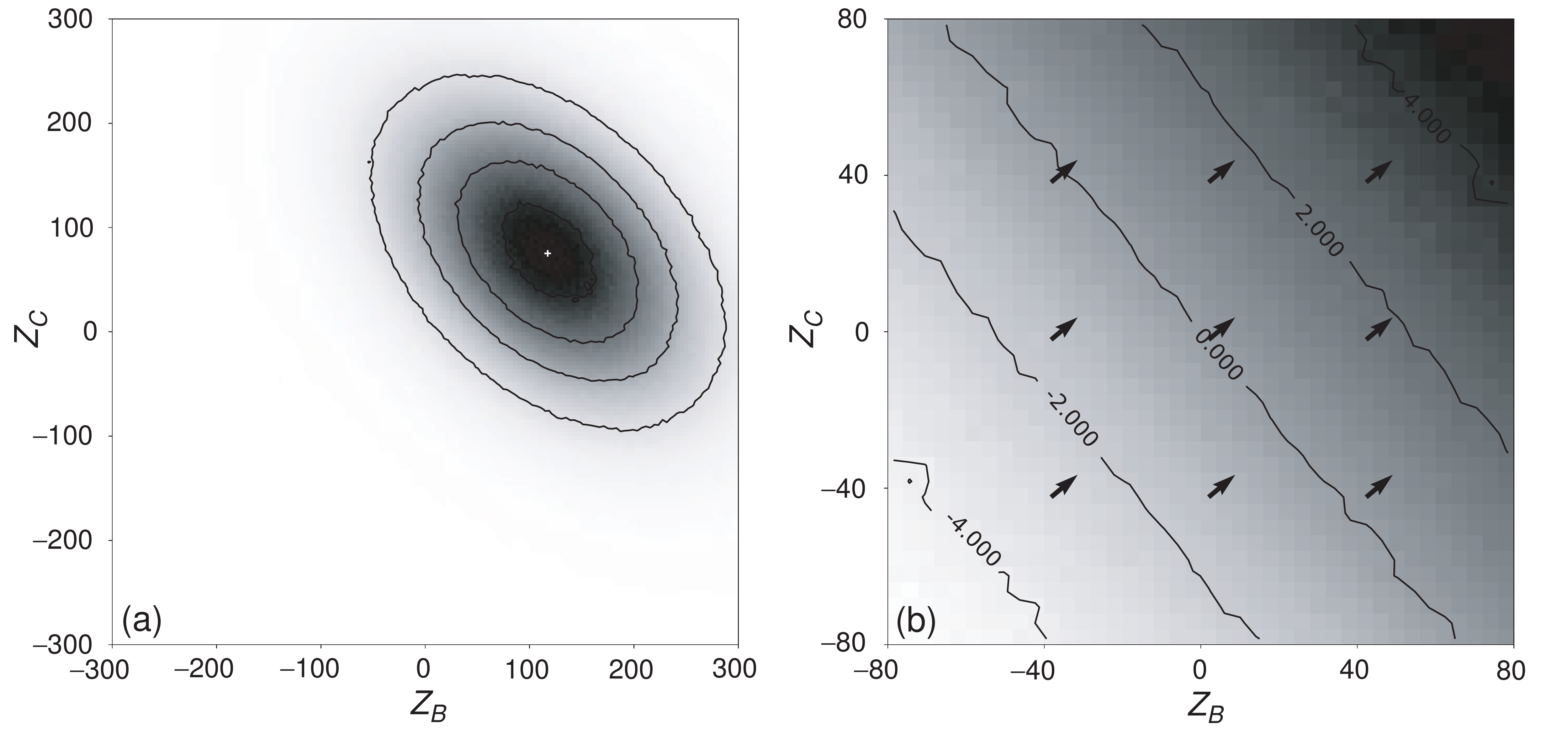}}
\end{minipage}
\begin{minipage}[t]{0.9\hsize}
\resizebox{1.0\hsize}{!}{\includegraphics{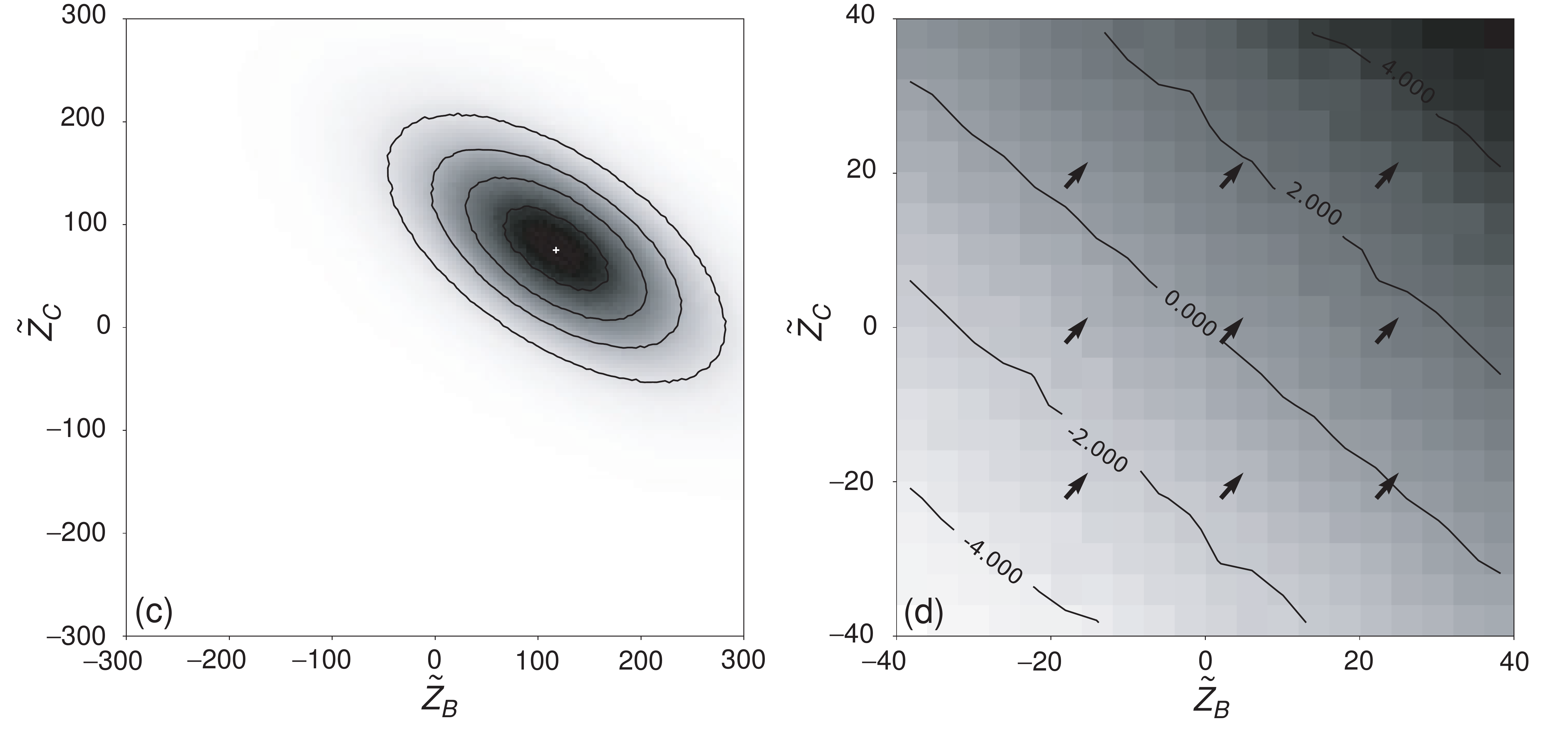}}
\end{minipage}
\caption{(a) Joint probability distribution $P(Z_C,Z_B,t)$ of the transferred charges $Z_C$ and $Z_B$ at time $t=20$. The center of this distribution marked with the symbol + corresponds to the mean values $\langle Z_B\rangle=117.43$ and $\langle Z_C\rangle=75.21$. Several contours of the distribution are also plotted. (b)~The function $\ln\left[P(Z_C,Z_B,t)/P(-Z_C,-Z_B,t)\right]$ versus $Z_C$ and $Z_B$ at the same time $t=20$. Several contours are shown. The arrows indicate the gradient of the distribution.  The finite-time affinities take the values $A_B(t=20)=0.0387$ and $A_C(t=20)=0.0326$. (c)~Joint probability distribution $P(\tilde Z_C,\tilde Z_B,t)$ of the transferred total charges $\tilde Z_C$ and $\tilde Z_B$ including the displacement currents, at the same time $t=20$. This distribution is centered on the same mean values $\langle \tilde Z_B\rangle=117.43$ and $\langle \tilde Z_C\rangle=75.21$. (d)~The corresponding function $\ln\left[P(\tilde Z_C,\tilde Z_B,t)/P(-\tilde Z_C,-\tilde Z_B,t)\right]$ versus $\tilde Z_C$ and $\tilde Z_B$ at the same time $t=20$, giving the finite-time affinities $\tilde A_B(t=20)=0.0659$ and $\tilde A_C(t=20)=0.0752$.  For both cases, the affinities are set in the simulation to the value $A_C=A_B=0.1$. The simulation is carried out with the time step $dt=0.1$ and the statistics over $3\times 10^7$ trajectories. The pixels in the four panels are all of size $4\times 4$.}
\label{fig4}
\end{figure*}

The direct test of the fluctuation relation~(\ref{FT}) requires the availability of an overlap between the probability distributions $P(Z_C,Z_B,t)$ and $P(-Z_C,-Z_B,t)$.  Since the maxima of these distributions move apart under nonequilibrium conditions, the overlap rapidly decreases as time increases.  Therefore, the direct test of the fluctuation relation is restricted to short times. Nevertheless, the test is possible as shown in Fig.~\ref{fig4} for the joint probability distributions of the accumulated charge numbers without and with the displacement currents using the set of parameter values given in Table~\ref{tab_parameters_2}.  For the bare charge numbers, Fig.~\ref{fig4}(a) depicts the joint distribution itself at time $t=20$, which is roughly Gaussian and shifted with respect to the origin because of the elapsed time.  There is a significant overlap with the opposite distribution $P(-Z_C,-Z_B,t)$ and Fig.~\ref{fig4}(b) shows several contours of the two-dimensional function $\ln\left[P(Z_C,Z_B,t)/P(-Z_C,-Z_B,t)\right]$ in the plane of the variables $Z_C$ and $Z_B$.  These contours appear straight given the presence of statistical errors, in agreement with the prediction of the fluctuation theorem that the function should be linear.  The function $\ln\left[P(Z_C,Z_B,t)/P(-Z_C,-Z_B,t)\right]$ can thus be fitted to a linear function $A_C(t)\,Z_C+A_B(t)\,Z_B$, defining the finite-time affinities $A_C(t)$ and $A_B(t)$. However, their values remain smaller than the applied affinities $A_C=A_B=0.1$ because convergence is expected for $t\gg t_{\rm diff}$ and has not yet been reached in Fig.~\ref{fig4}.  

\begin{table*}[h]
\caption{The set of parameter values used in Sec.~\ref{sec:FT} and Sec.~\ref{sec:resp}.}
\begin{tabular}{>{\centering\arraybackslash}m{5cm}>{\centering\arraybackslash}m{3cm}||>{\centering\arraybackslash}m{4.5cm}>{\centering\arraybackslash}m{3cm}}
\hline
\hline
parameter & value & parameter & value \bigstrut \\ \hline
volume of each cell   & $\Omega=1000$ & section areas & $\Sigma=10000$, $\Sigma_B=5000$ \bigstrut \\ \hline
number of electrons for the \textit{Collector} & $\bar{N}_C=10000$ & number of holes for the \textit{Collector} & $\bar{P}_C=100$ \bigstrut \\ \hline
number of electrons for the \textit{Base} & $\bar{N}_B=100$ & number of holes for the \textit{Base} & $\bar{P}_B=10000$ \bigstrut \\ \hline
number of electrons for the \textit{Emitter}& $\bar{N}_E=10000$ & number of holes for the \textit{Emitter} & $\bar{P}_E=100$ \bigstrut \\ \hline
\hline
\end{tabular}
\label{tab_parameters_2}
\end{table*}

\begin{figure*}
\begin{minipage}[h]{0.8\hsize}
\resizebox{0.65\hsize}{!}{\includegraphics{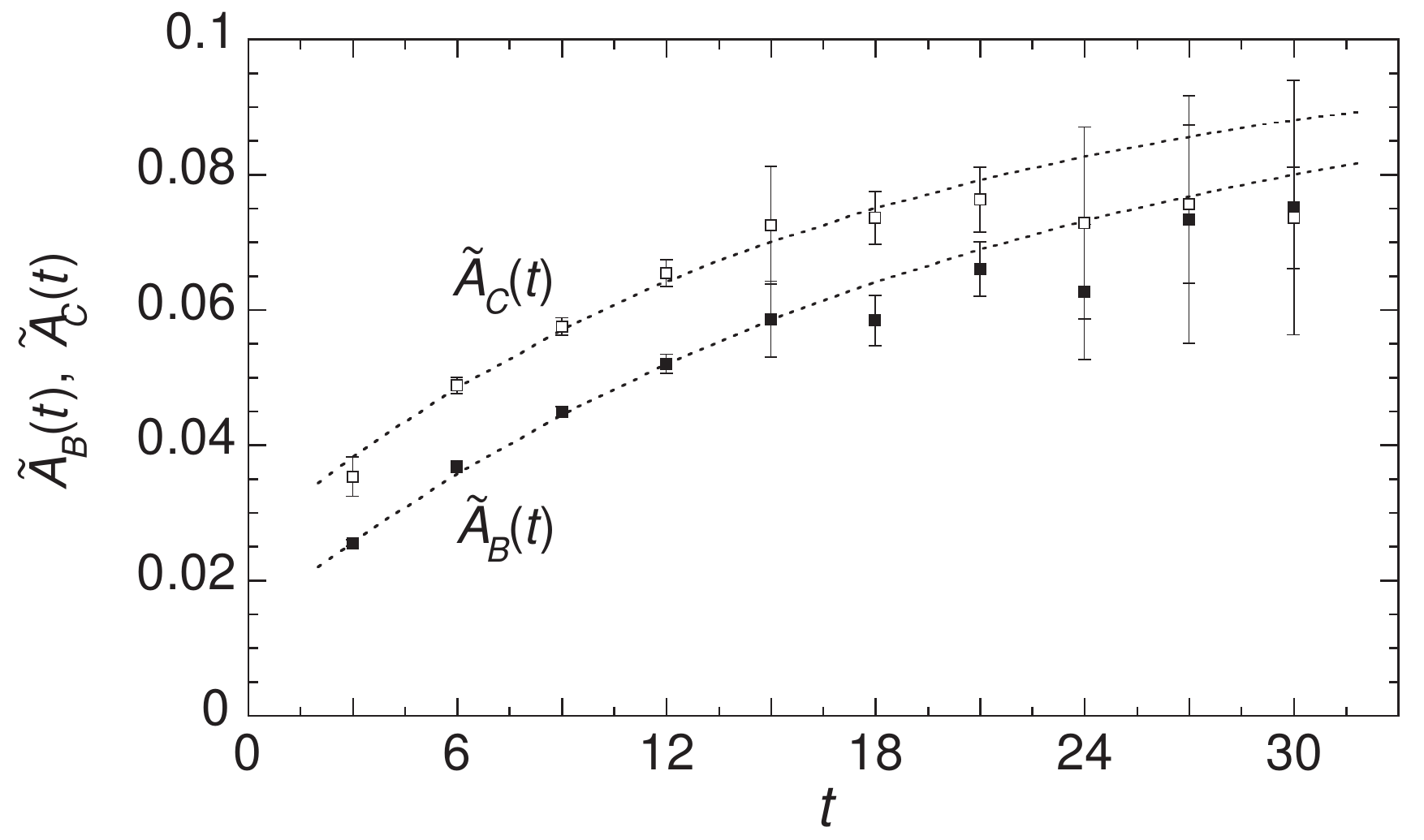}}
\end{minipage}
\caption{The finite-time affinities $\tilde A_C(t)$ and $\tilde A_B(t)$ versus time $t$ in the same conditions as in Fig.~\ref{fig4}(c) and Fig.~\ref{fig4}(d) for the transferred total charges $\tilde Z_C$ and $\tilde Z_B$ including the displacement currents.  These affinities are obtained by fitting $\ln\left[P(\tilde Z_C,\tilde Z_B,t)/P(-\tilde Z_C,-\tilde Z_B,t)\right]$ to the linear function $\tilde A_C(t)\,\tilde Z_C+\tilde A_B(t)\,\tilde Z_B$.   The dashed lines show the fits $\tilde A_C(t)\simeq 0.1-0.074\times\exp(-t/16.52)$ and $\tilde A_B(t)\simeq 0.1-0.086\times\exp(-t/20.61)$.}
\label{fig5}
\end{figure*}

\begin{table}[h]
\caption{The comparison between the numerical affinities and their theoretical expectations. The statistics used to evaluate the numerical affinities is obtained by simulations with the time step $dt=0.05$, the total time $t=2.5\times 10^3$, and $5\times 10^5$ trajectories for every case.}
\vskip 0.3 cm
\begin{tabular}{rrrrr}
\hline
\hline
case & $A_C^{\rm (th)}$ & $A_C^{\rm (num)}\qquad$ & $A_B^{\rm (th)}$ & $A_B^{\rm (num)}\qquad$ \bigstrut \\ \hline
1  & $1.0$ & $0.9914\pm 0.0034$ & $0.7$ & $0.6942\pm 0.0027$ \bigstrut \\ 
2  & $0.8$ & $0.7919\pm 0.0025$ & $0.4$ & $0.3952\pm 0.0019$ \bigstrut \\ 
3  & $0.5$ & $0.5018\pm 0.0033$ & $1.2$ & $1.2007\pm 0.0041$ \bigstrut \\ 
4 & $0.0$ & $0.0000\pm 0.0000$ & $0.0$ & $0.0000\pm 0.0000$ \bigstrut \\ 
5  & $-0.4$ & $-0.4002\pm 0.0018$ & $0.6$ & $0.5975\pm 0.0020$ \bigstrut \\ 
6  & $-0.5$ & $-0.4864\pm 0.0029$ & $-0.7$ & $-0.6864\pm 0.0029$ \bigstrut \\ 
7  & $-1.0$ & $-1.0058\pm 0.0039$ & $0.4$ & $0.4022\pm 0.0028$ \bigstrut \\ 
8  & $-1.2$ & $-1.3924\pm 0.0084$ & $1.3$ & $1.4118\pm 0.0084$ \bigstrut \\ 
\hline
\hline
\end{tabular}
\label{tab_cases}
\end{table}

As shown in Fig.~\ref{fig4}(c) and Fig.~\ref{fig4}(d), similar results hold for the joint probability distribution $P(\tilde Z_C,\tilde Z_B,t)$ of the charge numbers with the displacement currents.  As seen in Fig.~\ref{fig4}(c), the displacement currents have for effect that the distribution $P(\tilde Z_C,\tilde Z_B,t)$ is narrower than $P(Z_C,Z_B,t)$ depicted in Fig.~\ref{fig4}(a).  Consequently, the finite-time affinities $\tilde A_C(t)$ and $\tilde A_B(t)$ are larger than $A_C(t)$ and $A_B(t)$ and the convergence in time towards the asymptotic values of the affinities should be faster for the statistics of the transferred total charges $\tilde Z_C$ and $\tilde Z_B$ including the displacement currents, than for the statistics of the transferred charges $Z_C$ and $Z_B$.  Figure~\ref{fig5} confirms that the finite-time affinities $\tilde A_C(t)$ and $\tilde A_B(t)$ approach their asymptotic value $A_C=A_B=0.1$, as time increases.  Since the overlap between the opposite distributions rapidly decreases, statistical errors increase for $t>20$.  The exponential fits of the finite-time affinities provide estimations of the convergence times in the range of values expected by charge carrier diffusion.

In order to test the convergence of the finite-time affinities towards their asymptotic values over longer time scales, we develop a method using the following coarse-grained model,
\be
\begin{array}{c}
\textit{Collector}\autorightleftharpoons{$\scriptstyle W_{CE}$}{$\scriptstyle W_{EC}$}\textit{Emitter} \text{,}\\ 
\textit{Base}\autorightleftharpoons{$\scriptstyle W_{BE}$}{$\scriptstyle W_{EB}$}\textit{Emitter} \text{,}\\
\textit{Collector}\autorightleftharpoons{$\scriptstyle W_{CB}$}{$\scriptstyle W_{BC}$}\textit{Base} \text{,}
\end{array}
\label{model_CBE}
\ee
where the charges are supposed to be transferred between the three reservoirs with the transition rates $\{W_{kl}\}_{k,l=C,B,E}$, as formulated in Appendix~\ref{App:coarse}.  This constitutes the minimal model in the sense that the values of its rates can be fully determined from the knowledge of the mean currents and diffusivities, if the conditions of local detailed balance are satisfied.  This simple model is related to the Ebers-Moll transport model of bipolar junction transistors \cite{EM54,SS04}.   Given the values $J_C$, $J_B$, $D_{CC}$, $D_{BB}$, and $D_{CB}$ of the mean currents and the diffusivities, the six rates $W_{kl}$ can be determined, giving the values of the affinities according to $A_{kl}=\ln(W_{kl}/W_{lk})$ with $k,l=C,B,E$.  Since this model results from the coarse graining of the complete description, it has a domain of validity limited to moderate values of the applied voltages.  In this domain, the parameter values of the model can thus be fitted to the numerical values of the mean currents~(\ref{J_C})-(\ref{J_B}) and the diffusivities~(\ref{D_CC})-(\ref{D_CB}) of the full model in order to obtain the affinities.

Table~\ref{tab_cases} shows the comparison between the numerical affinities and the theoretical predictions for several cases.  Accurate agreement is found if the affinities remain moderate, confirming the convergence of the finite-time affinities $A_C(t)$ and $A_B(t)$ towards their expected asymptotic values~(\ref{eq_theoretical_affinity_C}) and~(\ref{eq_theoretical_affinity_B}) within the domain of validity of the model~(\ref{model_CBE}).  Despite the limited scope of application of this method, the agreement between the numerical and theoretical values of the affinities brings further numerical support to the fluctuation relation for the currents. In the next section, the consequences of the fluctuation theorem on the linear and nonlinear transport properties will be tested.

\section{Linear and nonlinear response properties}
\label{sec:resp}

\subsection{Deduction of the properties from the fluctuation theorem}

The fluctuation theorem provides a unified framework for deducing the Onsager reciprocal relations and their generalizations to the nonlinear transport properties \cite{S92,AG04,AG07JSM,HPPG11,BG18}.   For this purpose, it is convenient to introduce the cumulant generating function
\begin{align}
Q({\boldsymbol\lambda};{\bf A})\equiv\lim_{t\to\infty}-\frac{1}{t}\ln\int P_{A_C,A_B}(Z_C,Z_B,t)\, {\rm e}^{-\lambda_CZ_C-\lambda_BZ_B}\, dZ_C\, dZ_B \text{,} \label{CGF}
\end{align}
where ${\boldsymbol\lambda}=(\lambda_C,\lambda_B)$ are the so-called counting parameters and the macroscopic affinities are written in vectorial notation ${\bf A}=(A_C,A_B)$. As a consequence of the fluctuation theorem~(\ref{FT}), the cumulant generating function obeys the following symmetry relation
\begin{align}
Q({\boldsymbol\lambda};{\bf A})=Q({\bf A}-{\boldsymbol\lambda};{\bf A}) \text{.} \label{eq_symmetric_relation_of_FT}
\end{align} 
Now, the mean currents and the diffusivities can be obtained by taking the successive derivatives of the generating function~(\ref{CGF}) with respect to the counting parameters:
\begin{align}
& J_{\alpha}({\bf A})=\left.\frac{\partial Q({\boldsymbol\lambda};{\bf A})}{\partial\lambda_{\alpha}}\right\vert_{{\boldsymbol\lambda}={\bf 0}} \text{,} \\
& D_{\alpha\beta}({\bf A})=-\frac{1}{2}\left.\frac{\partial^2Q({\boldsymbol\lambda};{\bf A})}{\partial\lambda_{\alpha}\partial\lambda_{\beta}}\right\vert_{{\boldsymbol\lambda}={\bf 0}} \text{,} \label{dfn-D}
\end{align}
for $\alpha,\beta=C,B$. Besides, we may expand the mean currents in power series of the affinities as
\begin{align}
J_{\alpha}=\sum_{\beta}L_{\alpha,\beta}A_{\beta}+\frac{1}{2}\sum_{\beta,\gamma}M_{\alpha,\beta\gamma}A_{\beta}A_{\gamma}+\cdots 
\end{align}
in terms of the response coefficients defined by
\begin{align}
& L_{\alpha,\beta}=\left.\frac{\partial J_{\alpha}}{\partial A_{\beta}}\right\vert_{{\bf A}={\bf 0}}=\left.\frac{\partial^2 Q({\boldsymbol\lambda};{\bf A})}{\partial\lambda_{\alpha}\partial A_{\beta}}\right\vert_{{\boldsymbol\lambda}={\bf A}={\bf 0}} \text{,} \\
& M_{\alpha,\beta\gamma}=\left.\frac{\partial^2 J_{\alpha}}{\partial A_{\beta}\partial A_{\gamma}}\right\vert_{{\bf A}={\bf 0}}=\left.\frac{\partial^3 Q({\boldsymbol\lambda};{\bf A})}{\partial\lambda_{\alpha}\partial A_{\beta}\partial A_{\gamma}}\right\vert_{{\boldsymbol\lambda}={\bf A}={\bf 0}} \text{.}
\end{align}
The coefficients $L_{\alpha,\beta}$ characterize the linear response properties and the coefficients $M_{\alpha,\beta\gamma}$ the nonlinear response properties of the currents at second order in the affinities.  The coefficients of higher orders can also be introduced \cite{AG07JSM,BG18}.

If we take the derivatives of the symmetry relation Eq. (\ref{eq_symmetric_relation_of_FT}) with respect to $\lambda_\alpha$ and $A_\beta$, and set ${\boldsymbol\lambda}={\bf 0}$ and ${\bf A}={\bf 0}$, we obtain the fluctuation-dissipation relations
\begin{align}
L_{\alpha,\beta}= D_{\alpha\beta}({\bf A}={\bf 0})
\label{FDR}
\end{align}
and the Onsager reciprocal relations
\begin{align}
L_{\alpha,\beta}=L_{\beta,\alpha} \text{.} \label{ORR}
\end{align}
as a consequence of the symmetry $D_{\alpha\beta}=D_{\beta\alpha}$ resulting from the definition~(\ref{dfn-D}) of the diffusivities.

If we take a further derivative of the symmetry relation~(\ref{eq_symmetric_relation_of_FT}) with respect to $A_\gamma$ before setting ${\boldsymbol\lambda}={\bf 0}$ and ${\bf A}={\bf 0}$, we find that
\begin{align}
M_{\alpha,\beta\gamma}=\left(\frac{\partial D_{\alpha\beta}}{\partial A_{\gamma}}+\frac{\partial D_{\alpha\gamma}}{\partial A_{\beta}}\right)_{{\bf A}={\bf 0}} \text{,}
\label{M-dDdA}
\end{align}
giving the nonlinear response coefficient $M_{\alpha,\beta\gamma}$ in terms of the first responses of the diffusivities around equilibrium.  The relations~(\ref{M-dDdA}) as well as the Onsager reciprocal relations~(\ref{ORR}) find their origin in the microreversibility underlying the fluctuation theorem for currents \cite{S92,AGMT09,EHM09,CHT11,S12,G13}.

\begin{figure*}[h]
\begin{minipage}[t]{0.99\hsize}
\resizebox{1.0\hsize}{!}{\includegraphics{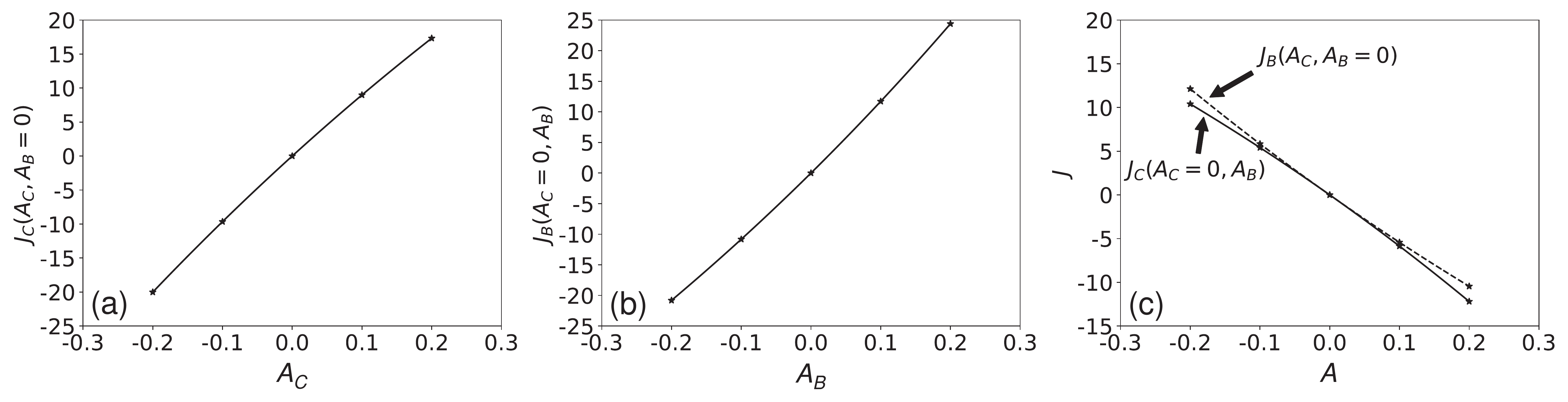}}
\end{minipage}
\caption{Mean charge currents versus one affinity with the other being zero: (a) The {\it Collector} current $J_C$ versus the {\it Collector} affinity $A_C$; (b) the {\it Base} current $J_B$ versus the {\it Base} affinity $A_B$; (c) the {\it Collector} (solid line) and {\it Base} (dashed line) currents versus the affinity of the other reservoir.  The asterisks are the numerical data from the simulation. The lines show the polynomials obtained from Lagrange interpolations using the data points. From the functions that are given by Lagrange polynomials, the first partial derivatives around the equilibrium point $(A_C=0,A_B=0)$ can be estimated, with the approximate values given in Table~\ref{tab_lin}.  The root mean squares on the data points are evaluated to be $\sigma_{J_C}\simeq 0.0020$ and $\sigma_{J_B}\simeq 0.0021$.  The simulations were carried out with the time step $dt=0.05$ and $10^9$ iterates for every data point.}
\label{fig6}
\end{figure*}

\begin{table*}[h]
\caption{The numerical values of the quantities used in the fluctuation-dissipation and the Onsager reciprocal relations.}
\vskip 0.3 cm
\begin{tabular}{|>{\centering\arraybackslash}m{4cm}|>{\centering\arraybackslash}m{4cm}||>{\centering\arraybackslash}m{3cm}|}
\hline
$L_{\alpha,\beta}$ & $\left.D_{\alpha\beta}\right\vert_{(0,0)}$  & $L_{\alpha,\beta}-\left.D_{\alpha\beta}\right\vert_{(0,0)}$ \bigstrut \\ \hline
$\left.\frac{\partial J_C}{\partial A_C}\right\vert_{(0,0)}=93.106\pm 0.019$ & $\left.D_{CC}\right\vert_{(0,0)}=92.991\pm 1.039$ & $0.115$ \bigstrut \\ \hline
$\left.\frac{\partial J_C}{\partial A_B}\right\vert_{(0,0)}=-56.288\pm 0.019$ & $\left.D_{CB}\right\vert_{(0,0)}=-56.343\pm 0.488$  & $0.055$ \bigstrut \\ \hline
$\left.\frac{\partial J_B}{\partial A_C}\right\vert_{(0,0)}=-56.303\pm 0.020$ & $\left.D_{BC}\right\vert_{(0,0)}=-56.343\pm 0.488$ &  $0.040$ \bigstrut \\ \hline
$\left.\frac{\partial J_B}{\partial A_B}\right\vert_{(0,0)}=112.603\pm 0.020$ & $\left.D_{BB}\right\vert_{(0,0)}=113.158\pm 0.487$ & $-0.555$ \bigstrut \\ \hline
\end{tabular}
\label{tab_lin}
\end{table*}

\begin{figure*}
\begin{minipage}[t]{0.99\hsize}
\resizebox{1.0\hsize}{!}{\includegraphics{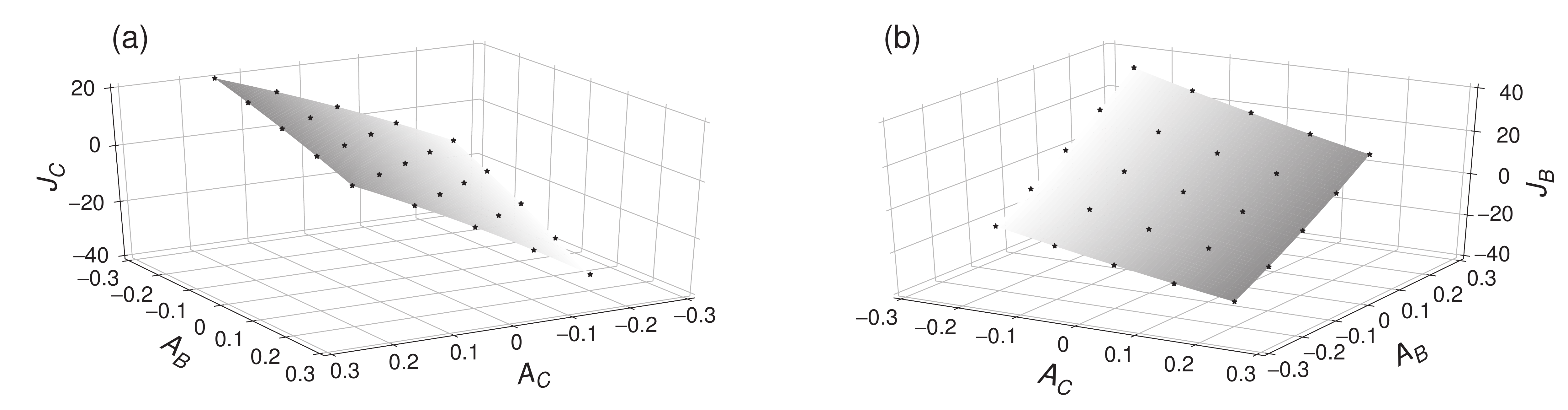}}
\end{minipage}
\caption{The mean charge currents as a function of the affinities $A_B$ and $A_C$: (a) The current $J_C$ from the \textit{Collector} to BJT; (b) The current $J_B$ from the \textit{Base} to BJT. The asterisks are the numerical data points from the simulation. The surfaces are obtained from Lagrange interpolation using the data points. Furthermore, the data points are used to get the second derivatives $\partial^2J_{\alpha}/\partial A_{\beta}\partial A_{\gamma}\vert_{(0,0)}$ around the equilibrium point $(A_C=0,A_B=0)$, as explained in Appendix~\ref{App:num}.  The numerical values of these second derivatives are given in Table~\ref{tab_nonlin}. The simulations were carried out with the time step $dt=0.05$ and $10^9$ iterates for every data point.}
\label{fig7}
\end{figure*}

\begin{figure*}
\begin{minipage}[t]{0.99\hsize}
\resizebox{1.0\hsize}{!}{\includegraphics{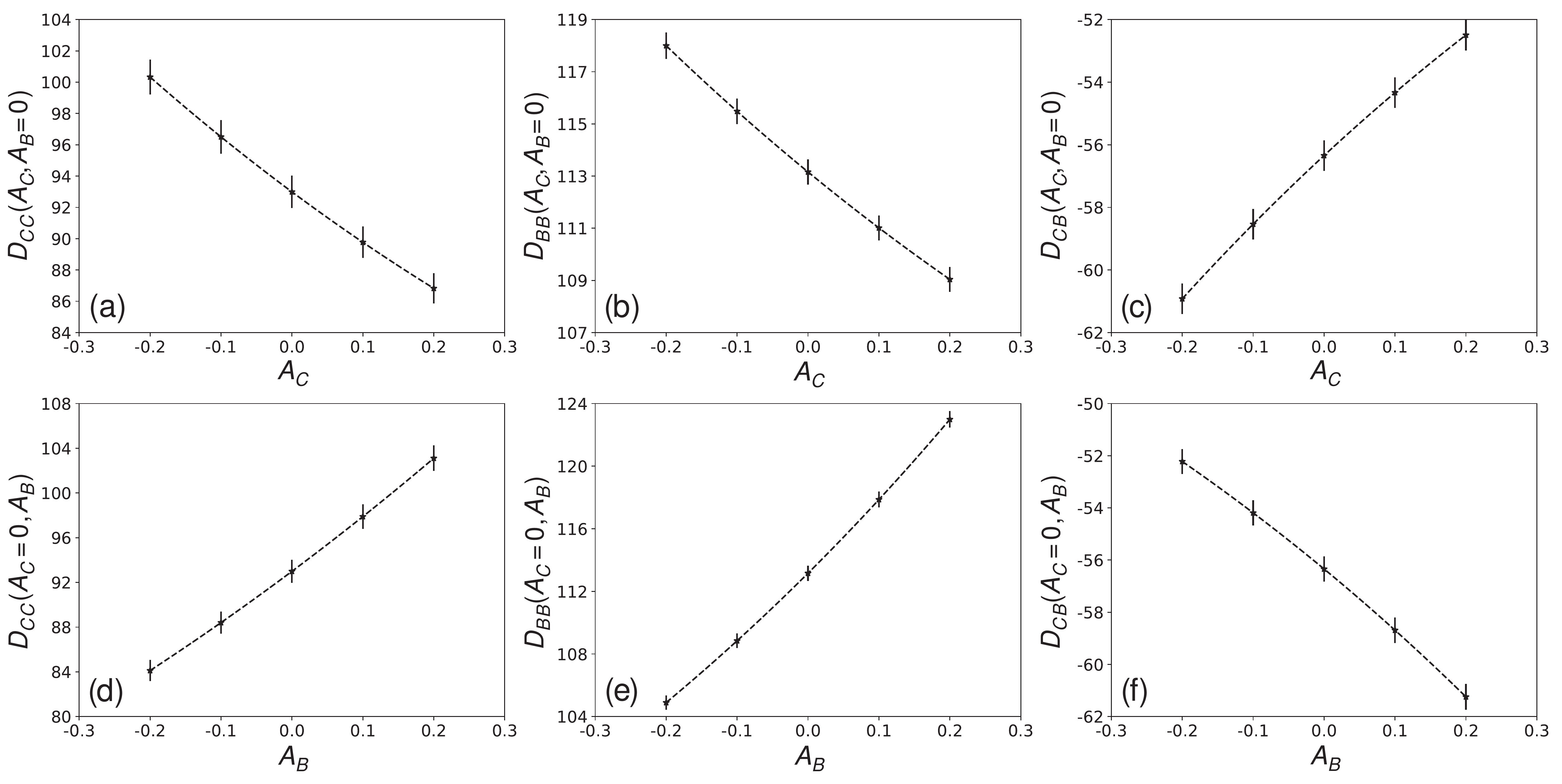}}
\end{minipage}
\caption{The diffusivities $D_{\alpha\beta}$ versus one affinity $A_{\gamma}$, the other affinity being set equal to zero. The numerical data points are plotted together with the error bars and the dashed lines give the Lagrange polynomial interpolations of the data points. These interpolations provide the first derivatives $\partial D_{\alpha\beta}/\partial A_{\gamma}\vert_{(0,0)}$ at the equilibrium point $(A_C=0,A_B=0)$. Their numerical values are given in Table~\ref{tab_nonlin}.  The simulations were carried out with the time step $dt=0.05$, the total time $t=2500$, and the statistics of $5\times 10^4$ trajectories for every data point.}
\label{fig8}
\end{figure*}

\begin{table*}
\caption{The numerical values of the quantities used in the nonlinear transport relations~(\ref{M-dDdA}).}
\vskip 0.3 cm
\begin{tabular}{|>{\centering\arraybackslash}m{4cm}|>{\centering\arraybackslash}m{4cm}|>{\centering\arraybackslash}m{4cm}||>{\centering\arraybackslash}m{3cm}|}
\hline
$M_{\alpha,\beta\gamma}$ & $R_{\alpha\beta,\gamma}$ & $R_{\alpha\gamma,\beta}$ & $M_{\alpha,\beta\gamma}-R_{\alpha\beta,\gamma}-R_{\alpha\gamma,\beta}$ \bigstrut \\ \hline
$\left.\frac{\partial^2 J_C}{\partial A_C^2}\right\vert_{(0,0)}=-67.388\pm 0.620$ & $\left.\frac{\partial D_{CC}}{\partial A_C}\right\vert_{(0,0)}=-33.642\pm 9.897$ & $\left.\frac{\partial D_{CC}}{\partial A_C}\right\vert_{(0,0)}=-33.642\pm 9.897$ & $-0.104$ \bigstrut \\ \hline
$\left.\frac{\partial^2 J_C}{\partial A_B^2}\right\vert_{(0,0)}=-45.325\pm 0.620$ & $\left.\frac{\partial D_{CB}}{\partial A_B}\right\vert_{(0,0)}=-22.474\pm 4.639$ & $\left.\frac{\partial D_{CB}}{\partial A_B}\right\vert_{(0,0)}=-22.474\pm 4.639$ & $-0.377$ \bigstrut \\ \hline
$\left.\frac{\partial^2 J_C}{\partial A_C\partial A_B}\right\vert_{(0,0)}=68.747\pm 0.097$ & $\left.\frac{\partial D_{CC}}{\partial A_B}\right\vert_{(0,0)}=47.409\pm 9.900$ & $\left.\frac{\partial D_{CB}}{\partial A_C}\right\vert_{(0,0)}=20.992\pm 4.642$ & $0.346$ \bigstrut \\ \hline
$\left.\frac{\partial^2 J_B}{\partial A_C^2}\right\vert_{(0,0)}=42.064\pm 0.667$ & $\left.\frac{\partial D_{CB}}{\partial A_C}\right\vert_{(0,0)}=20.992\pm 4.642$ & $\left.\frac{\partial D_{CB}}{\partial A_C}\right\vert_{(0,0)}=20.992\pm 4.642$ & $0.080$ \bigstrut \\ \hline
$\left.\frac{\partial^2 J_B}{\partial A_B^2}\right\vert_{(0,0)}=90.066\pm 0.665$ & $\left.\frac{\partial D_{BB}}{\partial A_B}\right\vert_{(0,0)}=45.068\pm 4.644$ & $\left.\frac{\partial D_{BB}}{\partial A_B}\right\vert_{(0,0)}=45.068\pm 4.644$ & $-0.070$ \bigstrut \\ \hline
$\left.\frac{\partial^2 J_B}{\partial A_C\partial A_B}\right\vert_{(0,0)}=-44.777\pm 0.107$ & $\left.\frac{\partial D_{CB}}{\partial A_B}\right\vert_{(0,0)}=-22.474\pm 4.639$ & $\left.\frac{\partial D_{BB}}{\partial A_C}\right\vert_{(0,0)}=-22.330\pm 4.630$ & $0.027$ \bigstrut \\ \hline
\end{tabular}
\label{tab_nonlin}
\end{table*}

\subsection{Numerical test of the linear transport properties}

In this subsection, we focus on the numerical test of the fluctuation-dissipation relations~(\ref{FDR}) and the Onsager reciprocal relation~(\ref{ORR}) for $\alpha,\beta=C,B$.  Here, we use the methods given in Appendix \ref{App:num} for the numerical evaluation of derivatives and their error analysis.

The evaluation of the linear response coefficients relies on the determination of the mean currents as a function of the affinities. To achieve this evaluation, we have computed the mean currents for several values of the affinities, as shown in Fig.~\ref{fig6}.  We have used the Lagrange interpolation method to obtain one-variable polynomials approximating $J_C(A_C,A_B= 0)$, $J_C(A_C= 0,A_B)$, $J_B(A_C,A_B= 0)$, and $J_B(A_C= 0,A_B)$ based on the numerical data plotted in Fig.~\ref{fig6}. Subsequently, the linear response coefficients can be computed by taking the first partial derivatives of the Lagrange polynomials at the equilibrium point $(A_C=0,A_B=0)$. Their numerical values are given in the first column of Table~\ref{tab_lin}.  This computation already confirms that the Onsager reciprocal relation $L_{C,B}=L_{B,C}$ is satisfied within the numerical accuracy.

Furthermore, the equilibrium values of the diffusivities are computed using Eqs.~(\ref{D_CC})-(\ref{D_CB}), giving the values in the second column of Table~\ref{tab_lin}.  The difference between the linear response coefficients and the diffusivities are reported in the third column of Table~\ref{tab_lin}, showing that the fluctuation-dissipation relations~(\ref{FDR}) are also satisfied within the numerical accuracy.

\subsection{Numerical test of the nonlinear transport properties}

The numerical values of the charge currents $J_C$ and $J_B$ are computed for different values of the affinities $A_C$ and $A_B$ in order to construct the two-variable functions $J_C(A_C,A_B)$ and $J_B(A_C,A_B)$ using  two-dimensional Lagrange interpolations, as shown in Fig.~\ref{fig7}. The values of second derivatives at the equilibrium point $(A_C=0,A_B=0)$,
\begin{align}
\left.\frac{\partial^2 J_{\alpha}}{\partial A_{\beta}\partial A_{\gamma}}\right\vert_{(0,0)} \qquad\mbox{for} \qquad \alpha,\,\beta,\,\gamma=C,\,B,
\end{align}
are thus numerically evaluated in order to determine the nonlinear response coefficients $M_{\alpha,\beta\gamma}$, using the numerical method explained in Appendix \ref{App:num}. On the other hand, the diffusivities $D_{\alpha\beta}$ are again computed using Eqs.~(\ref{D_CC})-(\ref{D_CB}), but for the transistor driven away from equilibrium.  They are plotted in Fig.~\ref{fig8} as functions of the affinities.  Therefore, the derivatives of the diffusivities with respect to the affinities
\begin{align}
R_{\alpha\beta,\gamma}\equiv\left.\frac{\partial D_{\alpha\beta}}{\partial A_{\gamma}}\right\vert_{(0,0)} 
\qquad\mbox{for} \qquad  \alpha,\,\beta,\,\gamma=C,\,B
\end{align}
can also be evaluated numerically at the equilibrium point $(A_C=0,A_B=0)$.  The results for the quantities $M_{\alpha,\beta\gamma}$ and $R_{\alpha\beta,\gamma}$ are given in Table \ref{tab_nonlin} where we calculate the differences, $M_{\alpha,\beta\gamma}-R_{\alpha\beta,\gamma}-R_{\alpha\gamma,\beta}$, testing the validity of the prediction~(\ref{M-dDdA}) of the fluctuation theorem beyond the linear transport properties.  We see that these differences are smaller than the numerical errors in agreement with the predictions.

\section{Conclusion and Perspectives}
\label{sec:conclude}

Using a spatially extended stochastic description of charge transport in bipolar $n$-$p$-$n$ junction transistors, we have shown in this paper that a fluctuation theorem holds for the two electric currents that are coupled together in the double junction of the transistor.  We have also shown that, as a corollary of the fluctuation theorem for the currents, nonlinear transport generalizations of the fluctuation-dissipation and Onsager reciprocal relations are satisfied in the transistor.  In particular, we have verified in detail that the second-order nonlinear response coefficients of the currents are related to the first-order responses of the diffusivities, as predicted by theory~\cite{AG04,AG07JSM,BG18}.

These results are based on stochastic partial differential equations describing the diffusion of electrons and holes, as well as their generation and recombination.  These stochastic diffusion-reaction equations are coupled to the Poisson equation for the electric potential and they obey local detailed balance.  The scheme is consistent with the laws of electricity, thermodynamics, and microreversibility.  The stochastic process is driven out of equilibrium by boundary conditions due to the voltages applied to the reservoirs in contact with the three ports of the transistor.  In this case, the transistor is the stage of a nonequilibrium steady state, manifesting highly nonlinear transport properties.  The key point raised in this paper is that, besides their amazing technological importance, transistors can be used to address the fundamental issue of microreversibility in nonequilibrium statistical physics.

The one-variable fluctuation theorem has already been experimentally investigated in linear $RC$ electric circuits \cite{GC05,JGC08}.  Our previous paper~\cite{GG18} has shown that the one-variable fluctuation theorem can be studied in nonlinear devices such as diodes.  In transistors, the experimental test of the two-variable fluctuation theorem can also be envisaged, either by the direct measurement of current fluctuations, or by testing its consequences, namely, the time-reversal symmetry relations generalizing the fluctuation-dissipation and Onsager reciprocal relations to the nonlinear transport properties.  Such tests would require accurate noise measurements with large enough statistics.  In this way, these symmetry relations, finding their origins in the fundamental law of microreversibility, could be tested experimentally in common devices of modern technology.

\vskip 1 cm

\section*{Acknowledgments}

The authors thank Sergio Ciliberto for stimulating discussions.
Financial support from the China Scholarship Council under the Grant No. 201606950037, the Universit\'e libre de Bruxelles (ULB), and the Fonds de la Recherche Scientifique - FNRS under the Grant PDR T.0094.16 for the project ``SYMSTATPHYS" is acknowledged.

\appendix

\begin{widetext}

\section{Discretized Markov jump process}
\label{App:Markov}

To describe the BJT by a Markov jump process, the system is spatially discretized into cells of volume $\Omega$, each containing some numbers $N_i$ and $P_i$ of electrons and holes, respectively.  These numbers are supposed to change in time because of random transitions at rates to be specified here below.  The Markov jump process is fully defined by these transition rates and the master equation ruling the time evolution of the probability that the cells contain given numbers of particles.  In the continuum limit, the Markov jump process leads to the stochastic reaction-diffusion equations~(\ref{eq-n})-(\ref{GWN-sig}), as shown in Appendix~\ref{App:Langevin}.  This method is similar to the one used in Refs.~\cite{AG09,GG18,G05}.

\subsection{Master equation of the process}
\label{master_eq}

At any time, the state of the discretized BJT is fully characterized by the electron numbers ${\bf N}=(N_i)_{i=1}^L$ and hole numbers ${\bf P}=(P_i)_{i=1}^L$ in all the cells.  The time evolution of these numbers is ruled by a Markov jump process corresponding to the following network:
\begin{equation}
\begin{array}{ccccccccccccc}
 & &  &  &  &  & \bar{N}_B & &  &  &  & \\
 & &  & & & & {\scriptstyle W_{mB}^{(+N)}}\downharpoonleft\upharpoonright{\scriptstyle W_{mB}^{(-N)}} & & & & & &  \\
\bar{N}_C & \autorightleftharpoons{$\scriptstyle W_0^{(+N)}$}{$\scriptstyle W_0^{(-N)}$} & N_1 & \cdots & N_{m-1} & \autorightleftharpoons{$\scriptstyle W_{m-1}^{(+N)}$}{$\scriptstyle W_{m-1}^{(-N)}$} & N_m & \autorightleftharpoons{$\scriptstyle W_{m}^{(+N)}$}{$\scriptstyle W_{m}^{(-N)}$} & N_{m+1} & \cdots & N_L & \autorightleftharpoons{$\scriptstyle W_L^{(+N)}$}{$\scriptstyle W_L^{(-N)}$} & \bar{N}_E \\
 & & {\scriptstyle W_1^{(+)}}\updownarrow{\scriptstyle W_1^{(-)}} & \cdots & {\scriptstyle W_{m-1}^{(+)}}\updownarrow{\scriptstyle W_{m-1}^{(-)}} &  & {\scriptstyle W_m^{(+)}}\updownarrow{\scriptstyle W_m^{(-)}} & &{\scriptstyle W_{m+1}^{(+)}}\updownarrow{\scriptstyle W_{m+1}^{(-)}} & \cdots & {\scriptstyle W_L^{(+)}}\updownarrow{\scriptstyle W_L^{(-)}} & &  \\
\bar{P}_C & \autorightleftharpoons{$\scriptstyle W_0^{(+P)}$}{$\scriptstyle W_0^{(-P)}$} & P_1 & \cdots & P_{m-1} & \autorightleftharpoons{$\scriptstyle W_{m-1}^{(+P)}$}{$\scriptstyle W_{m-1}^{(-P)}$} & P_m & \autorightleftharpoons{$\scriptstyle W_m^{(+P)}$}{$\scriptstyle W_m^{(-P)}$} & P_{m+1} & \cdots & P_L & \autorightleftharpoons{$\scriptstyle W_L^{(+P)}$}{$\scriptstyle W_L^{(-P)}$} & \bar{P}_E \\
 & & & & & & {\scriptstyle W_{mB}^{(+P)}}\upharpoonleft\downharpoonright{\scriptstyle W_{mB}^{(-P)}} & & & & & &  \\
&  &  &  &  &  & \bar{P}_B & & & & & &
\end{array} \nonumber
\end{equation}
On the left-hand side, the {\it Collector C} is a reservoir of electron and holes where their numbers $\bar{N}_C$ and $\bar{P}_C$ take fixed values.  On the right-hand side, it is the {\it Emitter E} that fixes the values of $\bar{N}_E$ and $\bar{P}_E$.  In the middle, similar transitions happen with the {\it Base~B}, fixing the values of $\bar{N}_B$ and $\bar{P}_B$.  The transitions with the rates $W_i^{(\pm N)}$ describe the diffusive transfers of electrons between the cells, and those with the rates $W_i^{(\pm P)}$ the diffusive transfers of holes.  The transitions with the rates $W_i^{(\pm)}$ describe the generation and recombination of electron-hole pairs, respectively.

The probability ${\cal P}(N_1, \dots, N_L, P_1, \dots, P_L,t)$ to find the system in a certain state is thus governed by the master equation
\begin{align}
\frac{d{\cal P}}{dt}=&\sum_{i=0}^{L} \Biggl[\left({\rm e}^{+\partial_{N_i}}{\rm e}^{-\partial_{N_{i+1}}}-1\right)W_i^{(+N)}{\cal P}+\left({\rm e}^{-\partial_{N_i}}{\rm e}^{+\partial_{N_{i+1}}}-1\right)W_i^{(-N)}{\cal P} +\left({\rm e}^{+\partial_{P_i}}{\rm e}^{-\partial_{P_{i+1}}}-1\right)W_i^{(+P)}{\cal P}+\left({\rm e}^{-\partial_{P_i}}{\rm e}^{+\partial_{P_{i+1}}}-1\right)W_i^{(-P)}{\cal P}\Biggr] \nonumber \\
+&\sum_{i=1}^{L}\Biggl[\left({\rm e}^{-\partial_{N_i}}{\rm e}^{-\partial_{P_i}}-1\right)W_i^{(+)}{\cal P}+\left({\rm e}^{+\partial_{N_i}}{\rm e}^{+\partial_{P_i}}-1\right)W_i^{(-)}{\cal P}\Biggr] \nonumber \\ 
+&\sum_{iB} \Biggl[\left({\rm e}^{-\partial_{N_i}}-1\right)W_{iB}^{(+N)}{\cal P}+\left({\rm e}^{+\partial_{N_i}}-1\right)W_{iB}^{(-N)}{\cal P} +\left({\rm e}^{-\partial_{P_i}}-1\right)W_{iB}^{(+P)}{\cal P}+\left({\rm e}^{+\partial_{P_i}}-1\right)W_{iB}^{(-P)}{\cal P}\Biggr] \, ,
\label{eq_master_equation}
\end{align}

\end{widetext}

\noindent with the transition rates given by
\begin{align}
& W_i^{(+N)}=\frac{D_n}{\Delta x^2}\psi(\Delta U_i^{(+N)})N_i \text{,} \\
& W_i^{(-N)}=\frac{D_n}{\Delta x^2}\psi(\Delta U_i^{(-N)})N_{i+1} \text{,} \\
& W_i^{(+P)}=\frac{D_p}{\Delta x^2}\psi(\Delta U_i^{(+P)})P_i \text{,} \\
& W_i^{(-P)}=\frac{D_p}{\Delta x^2}\psi(\Delta U_i^{(-P)})P_{i+1} \text{,}  \\
& W_i^{(+)}=\Omega k_+ \text{,} \\
& W_i^{(-)}=\Omega k_- \frac{N_i}{\Omega}\frac{P_i}{\Omega} \text{.}
\end{align}
For electron, the transition rates at the boundaries are given by
\begin{align}
& W_0^{(+N)}=\frac{D_n}{\Delta x^2}\psi(\Delta U_0^{(+N)})\bar{N}_C \text{,} \\
& W_0^{(-N)}=\frac{D_n}{\Delta x^2}\psi(\Delta U_0^{(-N)})N_1 \text{,} \\
& W_L^{(+N)}=\frac{D_n}{\Delta x^2}\psi(\Delta U_L^{(+N)})N_L \text{,} \\
& W_L^{(-N)}=\frac{D_n}{\Delta x^2}\psi(\Delta U_L^{(-N)})\bar{N}_E \text{,} \\
& W_{iB}^{(+N)}=\frac{D_n}{\Delta y^2}\psi(\Delta U_{iB}^{(+N)})\bar{N}_B \text{,} \\
& W_{iB}^{(-N)}=\frac{D_n}{\Delta y^2}\psi(\Delta U_{iB}^{(-N)})N_{i} \text{,}
\end{align}
and similar expressions for holes.  We note that, in the network shown above, the cell $i=m$ is the only one in contact with the {\it Base}, in which case the sum $\sum_{iB}$ in Eq.~(\ref{eq_master_equation}) has the sole term $i=m$.

$U$ is the total electrostatic energy stored in the BJT and $\Delta U$ is the energy difference associated with the change of the BJT state. $\psi(\Delta U)$ is a function defined by
\begin{align}
\psi(\Delta U)=\frac{\beta\Delta U}{\exp(\beta\Delta U)-1} \text{,}
\end{align}
which satisfies the local detailed balance condition
\begin{align}
\psi(\Delta U)=\psi(-\Delta U)\exp(-\beta\Delta U) \text{.}
\end{align}

\subsection{Discretized Poisson equation}
\label{App:Poisson}

The Poisson equation is replaced by its discretized version
\begin{align}
&\frac{\phi_{i+1}-2\phi_{i}+\phi_{i-1}}{\Delta x^{2}}+\frac{\phi_{B}-2\phi_{i}+\phi_{B}}{\Delta y^{2}}\chi_{iB}\nonumber\\ &=-\frac{e}{\epsilon\Omega}(P_{i}-N_{i}+D_{i}-A_{i})\hspace{1cm}(i=1,\dots,L) \text{,}
\end{align}
with the boundary conditions $\phi_0=\phi_C$ and $\phi_{L+1}=\phi_E$ at two ends of BJT, and the symbol $\chi_{iB}=1$ if the $i^{\rm th}$ cell is in contact with the {\it Base} and $\chi_{iB}=0$ otherwise. This linear system should be solved after every electron or hole transfer between cells. We suppose that the electric potential $\phi_B$ of the {\it Base} is set on both sides of the chain in the transverse $y$-direction, in order to get a symmetric geometry.

The electrostatic energy is given by

\begin{align}
U=\frac{1}{2}\, \pmb{\phi} \cdot \boldsymbol{\mathsf C}\cdot \pmb{\phi}
\end{align}
where the electric potential
\begin{align}
\pmb{\phi}=(\phi_{1},\dots,\phi_{i},\dots,\phi_{I})
\end{align}
obeys the discretized Poisson equation
\begin{align}
\boldsymbol{\mathsf C}\cdot \pmb{\phi}={\bf Z} 
\end{align}
with the symmetric matrix
\begin{align}
({\boldsymbol{\mathsf C}})_{ij}=a \left( -\delta_{i+1,j}+2\delta_{i,j}-\delta_{i-1,j}\right) + 2b \, \chi_{iB}\, \delta_{i,j}
\end{align}
expressed in terms of the Kronecker symbol such that $\delta_{i,j}=1$ if $i=j$ and $\delta_{i,j}=0$ otherwise, the coefficients
\begin{align}
a=\frac{\epsilon\Omega}{\Delta x^2} \, , \qquad b=\frac{\epsilon\Omega}{\Delta y^2} \, ,
\end{align}
and
\begin{align}
{\bf Z}=& e(\dots,P_i-N_i+D_i-A_i,\dots)\nonumber\\
&+a\, (\phi_{\rm C}, 0, \dots, 0, \phi_{E})\nonumber\\
&+2b\, (0,\dots,0,\phi_{B},\dots,\phi_{B},0,\dots,0) \text{.}
\end{align}
The change of electrostatic energy during the transfer of an electron of charge $-e$ from the $i^{\rm th}$ to the $(i+1)^{\rm th}$ cell is given by
\begin{align}
\Delta U_i^{(+N)}=\frac{1}{2}\left({\bf Z'} \cdot {\boldsymbol{\mathsf C}} ^{-1} \cdot {\bf Z'} - {\bf Z} \cdot {\boldsymbol{\mathsf C}}^{-1} \cdot{\bf Z}\right),
\end{align}
where
\begin{align}
Z'_k = Z_k+e\delta_{k,i}-e\delta_{k,i+1} \, ,
\end{align}
so that
\begin{align}
\Delta U_i^{(+N)}=& -e(\phi_{i+1}-\phi_i)\nonumber\\
&+\frac{e^2}{2}\Big[\left(\boldsymbol{\mathsf C}^{-1}\right)_{i,i}-2\left({\boldsymbol{\mathsf C}}^{-1}\right)_{i,i+1}+\left({\boldsymbol{\mathsf C}}^{-1}\right)_{i+1,i+1}\Big] \text{.} 
\end{align}
A similar expression holds for hole transfers since they have the charge $+e$, 
\begin{align}
\Delta U_i^{(+P)}=& +e(\phi_{i+1}-\phi_i)\nonumber\\
&+\frac{e^2}{2}\Big[\left(\boldsymbol{\mathsf C}^{-1}\right)_{i,i}-2\left({\boldsymbol{\mathsf C}}^{-1}\right)_{i,i+1}+\left({\boldsymbol{\mathsf C}}^{-1}\right)_{i+1,i+1}\Big] \text{.}
\end{align}
For electron transfers at the boundary, we have
\begin{align}
& \Delta U_0^{(+N)}=-e(\phi_1-\phi_C)+\frac{e^2}{2}\left(\boldsymbol{\mathsf C}^{-1}\right)_{1,1} \text{,} \\
& \Delta U_0^{(-N)}=e(\phi_1-\phi_C)+\frac{e^2}{2}\left(\boldsymbol{\mathsf C}^{-1}\right)_{1,1} \text{,} \\
& \Delta U_L^{(+N)}=-e(\phi_E-\phi_L)+\frac{e^2}{2}\left(\boldsymbol{\mathsf C}^{-1}\right)_{L,L} \text{,} \\
& \Delta U_L^{(-N)}=e(\phi_E-\phi_L)+\frac{e^2}{2}\left(\boldsymbol{\mathsf C}^{-1}\right)_{L,L} \text{,} \\
& \Delta U_{iB}^{(+N)}=-e(\phi_i-\phi_B)+\frac{e^2}{2}\left(\boldsymbol{\mathsf C}^{-1}\right)_{i,i} \text{,} \\
& \Delta U_{iB}^{(-N)}=e(\phi_i-\phi_B)+\frac{e^2}{2}\left(\boldsymbol{\mathsf C}^{-1}\right)_{i,i} \text{,}
\end{align}
and similar expressions for holes.

\section{Langevin stochastic process}
\label{App:Langevin}

In the limit where $N_i\gg 1$ and $P_i\gg 1$, the Markov jump process described here above can be replaced by a Langevin stochastic process~\cite{GG18,G05}, which is ruled by another master equation obtained by expanding the operators $\exp(\pm\partial_X)$ up to second order in the partial derivatives $\partial_X$ in Eq.~(\ref{eq_master_equation}).  In this way, we find that the corresponding probability density~$\mathscr P$ obeys the following Fokker-Planck equation:
\begin{align}
\partial_t{\mathscr P}=&\sum_{i=1}^{L}\Bigg\{-\partial_{N_i}\left[\left(W_{i-1}^{(+N)}-W_{i-1}^{(-N)}-W_i^{(+N)}+W_i^{(-N)}\right){\mathscr P}\right] \nonumber\\
&+\partial_{N_i}^2\left[\frac{1}{2}\left(W_{i-1}^{(+N)}+W_{i-1}^{(-N)}+W_i^{(+N)}+W_i^{(-N)}\right){\mathscr P}\right] \nonumber\\
&+\partial_{N_i}\partial_{N_{i+1}}\left[-\left(W_i^{(+N)}+W_i^{(-N)}\right){\mathscr P} \right]+(N\rightleftharpoons P)\Bigg\}\nonumber\\
+&\sum_{i=1}^{L}\Bigg\{-\left(\partial_{N_i}+\partial_{P_i}\right)\left[\left(W_i^{(+)}-W_i^{(-)}\right){\mathscr P}\right]\nonumber\\
&+\left(\partial_{N_i}+\partial_{P_i}\right)^2\left[\frac{1}{2}\left(W_i^{(+)}+W_i^{(-)}\right){\mathscr P}\right]\Bigg\} \nonumber\\
+&\sum_{iB}\Bigg\{-\partial_{N_i}\left[\left(W_{iB}^{(+N)}-W_{iB}^{(-N)}\right){\mathscr P}\right]\nonumber\\
&+\partial_{N_i}^2\left[\frac{1}{2}\left(W_{iB}^{(+N)}+W_{iB}^{(-N)}\right){\mathscr P}\right] +(N\rightleftharpoons P)\Bigg\}\text{.}
\end{align}
This shows that the variables $N_i$ and $P_i$ obey stochastic differential equations of Langevin type:
\begin{align}
& \frac{dN_i}{dt}=F_{i-1}^{(N)}-F_{i}^{(N)}+R_i+\chi_{iB}F_{iB}^{(N)} \text{,} \\
& \frac{dP_i}{dt}=F_{i-1}^{(P)}-F_{i}^{(P)}+R_i+\chi_{iB}F_{iB}^{(P)} \text{,}
\end{align}
with the following fluxes and reaction rates:
\begin{align}
& F_i^{(N)}=W_i^{(+N)}-W_i^{(-N)}+\sqrt{W_i^{(+N)}+W_i^{(-N)}}\xi_i^{(N)}(t) \text{,} \\
& F_i^{(P)}=W_i^{(+P)}-W_i^{(-P)}+\sqrt{W_i^{(+P)}+W_i^{(-P)}}\xi_i^{(P)}(t) \text{,} \\
& R_i=W_i^{(+)}-W_i^{(-)}+\sqrt{W_i^{(+)}+W_i^{(-)}}\eta_i(t) \text{,} \\
& F_{iB}^{(N)}=W_{iB}^{(+N)}-W_{iB}^{(-N)}+\sqrt{W_{iB}^{(+N)}+W_{iB}^{(-N)}}\xi_{iB}^{(N)}(t) \text{,} \\
& F_{iB}^{(P)}=W_{iB}^{(+P)}-W_{iB}^{(-P)}+\sqrt{W_{iB}^{(+P)}+W_{iB}^{(-P)}}\xi_{iB}^{(P)}(t) \text{,}
\end{align}
expressed in terms of the Gaussian white noises:
\begin{align}
& \langle\xi_i^{(N)}(t)\rangle=\langle\xi_i^{(P)}(t)\rangle=\langle\eta_i(t)\rangle=\langle\xi_{iB}^{(N)}(t)\rangle=\langle\xi_{iB}^{(P)}(t)\rangle=0 \text{,} \\
& \langle\xi_i^{(N)}(t)\,\xi_j^{(N)}(t')\rangle=\delta_{i,j}\delta(t-t') \text{,} \\
& \langle\xi_i^{(P)}(t)\,\xi_j^{(P)}(t')\rangle=\delta_{i,j}\delta(t-t') \text{,} \\
& \langle\eta_i(t)\,\eta_j(t')\rangle=\delta_{i,j}\delta(t-t') \text{,} \\
& \langle\xi_{iB}^{(N)}(t)\,\xi_{iB}^{(N)}(t')\rangle=\delta(t-t') \text{,} \\
& \langle\xi_{iB}^{(P)}(t)\,\xi_{iB}^{(P)}(t')\rangle=\delta(t-t') \text{,} \\
& \langle\xi^{(N)}(t)\,\xi^{(P)}(t')\rangle=\langle\xi_{iB}^{(N)}(t)\,\xi_{iB}^{(P)}(t')\rangle=0 \text{,} \\
& \langle\eta(t)\,\xi(t')\rangle=\langle\eta(t)\,\xi_{iB}(t')\rangle=\langle\xi(t)\,\xi_{iB}(t')\rangle=0 \text{.}
\end{align}
These Langevin stochastic equations are numerically implemented by discretizing time into equal intervals $\Delta t$ and replacing the white noises by independent identically distributed Gaussian random variables.  The stochastic partial differential equations (\ref{eq-n})-(\ref{GWN-sig}) are recovered in the continuum limit \cite{GG18}.

\section{Coarse-grained Markov jump process}
\label{App:coarse}

For the simple coarse-grained model~(\ref{model_CBE}), the joint probability distribution $P(Z_C,Z_B,t)$ to observe the charge transfers $Z_C$ and $Z_B$ during the time interval $[0,\,t]$ evolves according to the following master equation
\begin{align}
&\frac{d}{dt}P(Z_C,Z_B,t) \nonumber\\
& =W_{CE}P(Z_C-1,Z_B,t)+W_{EC}P(Z_C+1,Z_B,t)\nonumber\\
& +W_{BE}P(Z_C,Z_B-1,t)+W_{EB}P(Z_C,Z_B+1,t)\nonumber\\
& +W_{CB}P(Z_C-1,Z_B+1,t)+W_{BC}P(Z_C+1,Z_B-1,t)\nonumber\\
& -\left(W_{CE}+W_{EC}+W_{BE}+W_{EB}+ W_{CB}+W_{BC}\right)P(Z_C,Z_B,t) \text{.}
\label{eq_general_master_equation}
\end{align}
According to the central limit theorem, the joint probability distribution $P(Z_C,Z_B,t)$ after a long enough time interval $[0,\,t]$ becomes Gaussian of the following form,
\begin{align}
P({\bf Z},t)\simeq \frac{1}{4\pi t \sqrt{\det{\boldsymbol{\mathsf D}}}}\exp\left[-\frac{1}{4t}\, ({\bf Z}-{\bf J}\, t)^{\rm T}\cdot{\boldsymbol{\mathsf D}}^{-1}\cdot({\bf Z}-{\bf J}\, t)\right] \text{,} \label{eq_Gaussian_distribution}
\end{align}
with the vectorial and matricial notations 
\begin{align}
{\bf Z}=
\begin{pmatrix}
Z_C \\
Z_B
\end{pmatrix}
\text{,}\hspace{1cm}
{\bf J}=
\begin{pmatrix}
J_C \\
J_B
\end{pmatrix}
\text{,}\hspace{1cm}
{\boldsymbol{\mathsf D}}=
\begin{pmatrix}
D_{CC} & D_{CB} \\
D_{CB} & D_{BB}
\end{pmatrix} ,
\end{align}
and $^{\rm T}$ denoting the transpose. The mean charge currents ${\bf J}$ and the diffusivities ${\boldsymbol{\mathsf D}}$ can be numerically evaluated through
\begin{align}
{\bf J}=\lim_{t\to\infty} \frac{1}{t}\, \left\langle{\bf Z}(t)\right\rangle \text{,}\quad {\boldsymbol{\mathsf D}}=\lim_{t\to\infty} \frac{1}{2t}\, \left\langle\left[{\bf Z}(t)-{\bf J}\, t\right]\cdot\left[{\bf Z}(t)-{\bf J}\, t\right]^{\rm T}\right\rangle ,
\end{align}
where $\langle\cdot\rangle$ denotes the statistical average over the data sample. 

For this model, the mean currents and the diffusivities can be expressed in terms of the transition rates of the master equation~(\ref{eq_general_master_equation}) according to the following relations:
\begin{align}
& W_{CE}-W_{EC}+W_{CB}-W_{BC}=J_C \text{,} \label{eq_nonlinear_1} \\
& W_{BE}-W_{EB}+W_{BC}-W_{CB}=J_B \text{,} \label{eq_nonlinear_2} \\
& W_{CE}+W_{EC}+W_{CB}+W_{BC}=2D_{CC} \text{,} \label{eq_nonlinear_3} \\
& W_{BE}+W_{EB}+W_{BC}+W_{CB}=2D_{BB} \text{,} \label{eq_nonlinear_4} \\
& W_{CB}+W_{BC}=-2D_{CB} \text{.} \label{eq_nonlinear_5}
\end{align}
By local detailed balance, the affinities are given by
\begin{align}
& A_{CE}=\ln\left(\frac{W_{CE}}{W_{EC}}\right) \text{,} \label{A_CE}\\
& A_{CB}=\ln\left(\frac{W_{CB}}{W_{BC}}\right) \text{,} \label{A_CB}\\
& A_{BE}=\ln\left(\frac{W_{BE}}{W_{EB}}\right) \text{.} \label{A_BE}
\end{align}
The natural condition
\begin{align}
A_{CB}+A_{BE}=A_{CE}
\end{align}
leads to
\begin{align}
W_{CB}W_{BE}W_{EC}=W_{BC}W_{EB}W_{CE} \text{.} \label{eq_nonlinear_6}
\end{align}
Eqs. (\ref{eq_nonlinear_1})-(\ref{eq_nonlinear_5}) and~(\ref{eq_nonlinear_6}) form a set of six nonlinear equations that can be solved numerically with the Newton-Raphson method to find the six transition rates $\{W_{kl}\}_{k,l=C,B,E}$.  Thereafter, the affinities are readily evaluated by Eqs.~(\ref{A_CE})-(\ref{A_BE}). Taking the \textit{Emitter} as the reference reservoir, we may more shortly write $A_{CE}$ as $A_C$, and $A_{BE}$ as $A_B$. 

We note that these considerations lead to the Ebers-Moll transport model of bipolar junction transistors \cite{EM54,SS04} if we assume that $W_{CB}=J_s/\beta_R$, $W_{EB}=J_s/\beta_F$, and $W_{EC}=J_s\exp(\beta e V_{BC})$, where $J_s$ is the reverse saturation current, $\beta_R$~the reverse common emitter current gain, and $\beta_F$ the forward common emitter current gain, in addition to the local detailed balance conditions $W_{BC}=W_{CB}\exp(\beta e V_{BC})$ and $W_{BE}=W_{EB}\exp(\beta e V_{BE})$ given by Eqs.~(\ref{A_CB}) and~(\ref{A_BE}).  The well-known expressions for the mean currents of this model (e.g., given Ref.~\cite{SS04} pp. 387-389) are thus recovered from Eqs.~(\ref{eq_nonlinear_1}) and~(\ref{eq_nonlinear_2}) by using Eq.~(\ref{eq_nonlinear_6}).

\section{Numerical differentiation and error analysis}
\label{App:num}

The differentiation can be approximated by numerical differences using several points \cite{AS72}. Given the values of the one-variable function $f(x)$ at the five equispaced points $-2h$, $-h$, $0$, $h$, $2h$, we have the following centered-difference formulae
\begin{align}
& f^{\prime}(0)\simeq\frac{-f(2h)+8f(h)-8f(-h)+f(-2h)}{12h} \text{,} \label{eq_first_derivative} \\
& f^{\prime\prime}(0)\simeq\frac{-f(+2h)+16f(+h)-30f(0)+16f(-h)-f(-2h)}{12h^2} \text{,} \label{eq_second_derivative}
\end{align}
respectively giving the first- and second-order derivatives up to numerical errors of $O(h^4)$. These two difference formulae can be obtained using the Lagrange polynomial
\begin{align}
L_n(x)=\sum_{i=0}^n\left[\prod_{j=0,j\neq i}^n\left(\frac{x-x_j}{x_i-x_j}\right)\right]f(x_i)
\end{align}
that interpolates the five points at $x_i=-2h,\,-h,\,0,\,h,\,2h$. Here, it is easy to obtain Lagrange polynomial corresponding to the two-variable function $f(x,y)$ using points distributed on a grid 
\begin{align}
L(x, y)=\sum_{i,j}\left[\prod_{m\neq i,n\neq j}\left(\frac{x-x_m}{x_i-x_m}\right)\left(\frac{y-y_n}{y_j-y_n}\right)\right]f(x_i,y_j) \text{.}
\end{align}
The mixed second derivative of $f(x,y)$ at the point $(0, 0)$ can be approximated by the midpoint formula
\begin{align}
\frac{\partial^2 f}{\partial x\partial y}(0, 0)\simeq\frac{f(h_1, h_2)-f(h_1, -h_2)-f(-h_1, h_2)+f(-h_1, -h_2)}{4h_1h_2} \text{,} \label{eq_mixed_second_derivative}
\end{align}
which is accurate up to $O(h_1^2h_2^2)$.

Apart from the numerical error itself, another source of errors comes from the statistical evaluation of the function at the different points. Suppose that the variances of the numerical values of the function are denoted as $\sigma^2\left[f(x_i)\right]$ and $\sigma^2\left[f(x_i,y_j)\right]$, then the mean square errors on the derivative (\ref{eq_first_derivative}) can be evaluated as
\begin{align}
\sigma^2\left[f^{\prime}(0)\right] \simeq \frac{1}{144h^2}&\Big\{\sigma^2\left[f(2h)\right]+64\sigma^2\left[f(h)\right]\nonumber\\&+64\sigma^2\left[f(-h)\right]+\sigma^2\left[f(-2h)\right]\Big\} \text{,} 
\end{align}
up to a correction of $O(h^8)$ coming from the error in the numerical differentiation. Similar expressions hold for the mean square errors on the other derivatives~(\ref{eq_second_derivative}) and~(\ref{eq_mixed_second_derivative}).

Given the random sample $\{X_1,\dots,X_n\}$ of size $n$ from a Gaussian distribution of mean value $\mu$ and variance $\sigma^2$, the sample average is defined as $\langle X\rangle=(1/n)\sum_{i=1}^nX_i$, having the expected value equal to $\mu$. The sample average $\langle X\rangle$ has the mean square error ${\rm MSE}\left(\langle X\rangle\right)=\sigma^2/n$.  The unbiased sample variance $S_{n-1}^2=\sum_{i=1}^n(X-\langle X\rangle)^2/(n-1)$ has the expected value $\sigma^2$ and its mean square error is equal to ${\rm MSE}\left(S_{n-1}^2\right)=2\sigma^4/(n-1)$.  If we define the average current $J=\langle X\rangle/t$ and diffusivity $D=S_{n-1}^2/(2t)$, their mean square errors can thus be estimated as
\begin{align}
{\rm MSE}\left(J\right)=\frac{\sigma^2}{nt^2}\simeq\frac{2D}{nt} \text{,}\quad {\rm MSE}\left(D\right)=\frac{2\sigma^4}{4t^2(n-1)}\simeq\frac{2D^2}{n-1} \text{.}
\end{align}

The procedure used to estimate the error on the numerical computation of the affinities $A_C=\ln\left(W_{CE}/W_{EC}\right)$ and $A_B=\ln\left(W_{BE}/W_{EB}\right)$ by the method of Appendix~\ref{App:coarse} is the following.  The expressions of the affinities are differentiated with respect to the mean values of the currents and diffusivities to obtain linear approximations such as
\begin{align}
\Delta A_C\simeq a\,\Delta J_C+b\,\Delta J_B+c\,\Delta D_{CC}+d\,\Delta D_{BB}+e\,\Delta D_{CB} \text{,}
\end{align}
in terms of some coefficients $a$, $b$, $c$, $d$, and $e$, which are related to the rates. Accordingly, the mean square error is estimated as
\begin{align}
\sigma^2(A_C)\simeq &a^2\sigma^2(J_C)+b^2\sigma^2(J_B)\nonumber\\
&+c^2\sigma^2(D_{CC})+d^2\sigma^2(D_{BB})+e^2\sigma^2(D_{CB}) \text{,}
\end{align}
and similarly for the error on $A_B$.



\begin{thebibliography}{99}

\bibitem{SST51} W. Shockley, M. Sparks, and G. K. Teal, Phys. Rev. {\bf 83}, 151 (1951).

\bibitem{EM54} J. J. Ebers and J. L. Moll, Proc. IRE {\bf 42}, 1761 (1954).

\bibitem{SS04} A. S. Sedra and K. C. Smith, {\it Microelectronic circuits}, 5th edition (Oxford University Press, New York, 2004).

\bibitem{CC05} J.-P. Collinge and C. A. Collinge, {\it Physics of semiconductor devices} (Kluwer Academic Publishers, New York, 2005).

\bibitem{B05} K. F. Brennan, {\it Introduction to semiconductor devices} (Cambridge University Press, Cambridge UK, 2005).

\bibitem{SN07} S. M. Sze and K. K. Ng, {\it Physics of semiconductor devices}, 3rd~edition (Wiley, Hobboken, 2007).

\bibitem{O31a} L. Onsager, Phys. Rev. {\bf 37}, 405 (1931).

\bibitem{O31b} L. Onsager, Phys. Rev. {\bf 38}, 2265 (1931).

\bibitem{C45} H. B. G. Casimir, Rev. Mod. Phys. {\bf 17}, 343 (1945).

\bibitem{S92} R. L. Stratonovich, {\it Nonlinear Nonequilibrium Thermodynamics I} ( Springer-Verlag, Berlin, 1992).

\bibitem{AG04} D. Andrieux and P. Gaspard, J. Chem. Phys. {\bf 121}, 6167 (2004).

\bibitem{AG07JSM} D. Andrieux and P. Gaspard, J. Stat. Mech.: Th. Exp., P02006 (2007).

\bibitem{HPPG11} P. I. Hurtado, C. P\'erez-Espigares, J. J. del Pozo, and P. L. Garrido, Proc. Natl. Acad. Sci. (USA) {\bf 108}, 7704 (2011).

\bibitem{BG18} M. Barbier and P. Gaspard, J. Phys. A: Math. Theor. {\bf 51}, 355001 (2018).

\bibitem{AG07JSP} D. Andrieux and P. Gaspard, J. Stat. Phys. {\bf 127}, 107 (2007).

\bibitem{AG09} D. Andrieux and P. Gaspard, J. Stat. Mech. P02057 (2009).

\bibitem{AGMT09} D.~Andrieux, P.~Gaspard, T.~Monnai, and S.~Tasaki, New J. Phys. {\bf 11}, 043014 (2009); {\it Erratum}, {\it ibid.} {\bf 11}, 109802 (2009).

\bibitem{EHM09} M.~Esposito, U.~Harbola, and S.~Mukamel, Rev. Mod. Phys. {\bf 81}, 1665 (2009).

\bibitem{CHT11} M. Campisi, P. H\"anggi, and P. Talkner, Rev. Mod. Phys. {\bf 83}, 771 (2011); {\it Erratum}, {\it ibid.} {\bf 83}, 1653 (2011).

\bibitem{S12} U. Seifert, Rep. Prog. Phys. {\bf 75}, 126001 (2012).

\bibitem{G13} P. Gaspard, New J. Phys. {\bf 15}, 115014 (2013).

\bibitem{GG18} J. Gu and P. Gaspard, Phys. Rev. E {\bf 97}, 052138 (2018).

\bibitem{G76} D. T. Gillespie, J. Comput. Phys. {\bf 22}, 403 (1976).

\bibitem{BB00} Ya. M. Blanter and M. B\"uttiker, Phys. Rep. {\bf 336}, 1 (2000).

\bibitem{S38} W. Shockley, J. Appl. Phys. {\bf 9}, 635 (1938).

\bibitem{R39} S. Ramo, Proc. IRE {\bf 27}, 584 (1939).

\bibitem{S76} J. Schnakenberg, Rev. Mod. Phys. {\bf 48}, 571 (1976).

\bibitem{GGHK18} P. Gaspard, P. Grosfils, M.-J. Huang, and R. Kapral, J. Stat. Mech. 123206 (2018).

\bibitem{GC05} N. Garnier and S. Ciliberto, Phys. Rev. E {\bf 71}, 060101 (2005).

\bibitem{JGC08} S. Joubaud, N. B. Garnier, and S. Ciliberto, Europhys. Lett. {\bf 82} 30007 (2008).

\bibitem{G05} P. Gaspard, New J. Phys. {\bf 7}, 77 (2005).

\bibitem{AS72} M. Abramowitz and I. A. Stegun, {\it Handbook of mathematical functions} (Dover, New York, 1972).

\end{thebibliography}
\end{document}